\documentclass[iop,apj]{emulateapj} %,author-year
\usepackage[utf8]{inputenc}
\usepackage[T1]{fontenc}
\usepackage[english]{babel}
\usepackage{amsmath}
\usepackage[upint]{stix}
\usepackage{helvet}
\usepackage{graphicx}
\usepackage{color} %in order to have colored text

\newcommand{\td}[1]{\, \mbox{d} #1 \,}
\newcommand{\intl}{\int\limits}

\newcommand{\SF}[3]{\; H\left[ #1;\, #2,\, #3 \right]}  %Step-Function
\newcommand{\DF}[1]{\; \delta\left( #1 \right)}  %Delta-Function
\newcommand{\fermi}{\textit{Fermi}-LAT}
\newcommand{\g}{\ensuremath{\gamma}}
\newcommand{\cta}{\object{CTA\,102}} %\xspace
\newcommand{\zred}{z_{\rm red}}

\newcommand{\logb}[1]{\ln{\left( #1 \right)}}
\newcommand{\p}{^{\prime}}
\newcommand{\obs}{^{\rm obs}}
\newcommand{\est}[3]{\left( \frac{#1}{#2} \right)^{#3}}
\newcommand{\E}[1]{\times 10^{#1}}
\newcommand{\change}[1]{#1}

\begin{document}

\shorttitle{Cloud ablation in \cta}
\shortauthors{M. Zacharias, et al.}

\title{The extended flare in \cta\ in 2016 and 2017 within a hadronic model through cloud ablation by the relativistic jet}

\author{M. Zacharias$^{1,2}$, M. B\"ottcher$^1$, F. Jankowsky$^3$, J.-P. Lenain$^4$, S.~J. Wagner$^3$, A. Wierzcholska$^5$}
\email{mzacharias.phys@gmail.com}
\affiliation{$^1$Centre for Space Science, North-West University, Potchefstroom, 2520, South Africa \\
$^2$Insitut f\"ur theoretische Physik IV, Ruhr-Universit\"at Bochum, D-44780 Bochum, Germany \\
$^3$Landessternwarte, Universit\"at Heidelberg, K\"onigstuhl, D-69117 Heidelberg, Germany \\
$^4$Sorbonne Universit\'e, Universit\'e Paris Diderot, Sorbonne Paris Cit\'e, CNRS/IN2P3, Laboratoire de Physique Nucl\'eaire et de Hautes Energies, LPNHE, 4 Place Jussieu, F-75252 Paris, France \\
$^5$Institute of Nuclear Physics, Polish Academy of Sciences, PL-31342 Krakow, Poland
}

\begin{abstract}
The flat spectrum radio quasar \cta\ (redshift 1.037) exhibited a tremendously bright 4-months long outburst from late 2016 to early 2017. In a previous paper, we interpreted the event as the ablation of a gas cloud by the relativistic jet. The multiwavelength data have been reproduced very well within this model using a leptonic emission scenario. Here we expand that work by using a hadronic scenario, which gives us greater freedom with respect to the location of the emission region within the jet. This is important, since the inferred gas cloud parameters depend on the distance from the black hole. While the hadronic model faces the problem of invoking super-Eddington jet luminosities, it reproduces well the long-term trend and also days-long subflares. While the latter result in inferred cloud parameters that match those expected for clouds of the broad-line region, the long-term trend is not compatible with such an interpretation. We explore the possibilities that the cloud is from the atmosphere of a red giant star or comes from a star-forming region that passes through the jet. The latter could also explain the much longer-lasting activity phase of \cta\ from late 2015 till early 2018.
\end{abstract}


\maketitle
%\linenumbers
%
%################################################################################################################################
%
\section{Introduction}
%
%A particularly interesting case is the interaction of the jet with an obstacle, such as a star \citep{bk79,k94,pmlh14,b15}, its wind \citep{abr09,dcea17} or a gas cloud \citep{abr10,bpb12}. Most of these models have in common that the obstacle is already fully inside the jet before the start of the interaction. However, given the strong pressure of the relativistically moving matter of the jet, interactions will start as soon as the obstacle hits the jet, since the jet will look like a strong shock. Simulations of shock/cloud interactions have shown that a cloud will be quickly ripped apart \citep{kkc94,pfb02}. Recent simulations of a jet/cloud \citep{bpb12} or jet/star \citep{pbb17} interaction, where the penetration process is included, reveal that the obstacle is (partially) ablated, and a significant amount of matter is mixed into the jet flow.
%
%This is easy to see for a gas cloud, given that it is mainly confined by its own, rather weak gravity. The ram pressure of the jet will immediately start to ablate the outer layers of the cloud while it starts to penetrate the jet. The mass loss of the cloud will weaken its structural integrity even before it has fully penetrated the jet. As we will discuss below, the cloud will be ablated and carried along by the jet. Depending on the cloud parameters, such as size and velocity, this might lead to pronounced and prolonged jet activity, when the additional material in the jet reaches an internal shock located downstream of the cloud penetration site. 
%
\cta\ is a flat spectrum radio quasar (FSRQ), located half-way across the observable Universe at a redshift $\zred = 1.037$. As all FSRQ \citep{br74}, the relativistic jet of \cta\ is closely aligned with the line-of-sight and its spectral energy distribution (SED) exhibits the well-known double-humped structure. The low-energy hump is attributed to electron synchrotron emission, while the high energy hump is interpreted either as electron inverse-Compton (IC) emission or due to hadronic emission processes. The accretion disk luminosity in \cta\ is $L_{\rm disk}\p = 3.8\times 10^{46}\,$erg/s \citep{zcst14}. The mass of the central black hole is estimated at $M_{\rm bh}\sim 8.5 \times 10^8\, M_{\odot}$ \citep{zcst14} giving an Eddington luminosity of $L_{\rm Edd}\p\sim 1.1\times 10^{47}\,$erg/s. The luminosity of the broad-line region (BLR) is $L_{\rm BLR}\p = 4.14\times 10^{45}\,$erg/s with a radius of $R_{\rm BLR}\p = 6.7\times 10^{17}\,$cm \citep{pft05}. \cite{mea11} report a tentative detection of a dusty torus (DT) with a luminosity $L_{\rm DT}\p = 7.0\E{45}\,$erg/s. Scaling relations \citep[e.g.,][]{Hea12} then provide a radius of $R_{\rm DT}\p = 6.18\E{18}\,$cm$\,\sim 2\,$pc (all quantities given in the host galactic frame).
% JPL: in the *host* galactic frame, right ? 
% MZ: Yes, thanks.

\cta\ has been under continuous surveillance at high-energy $\gamma$-rays (HE, $E>100\,$MeV) since the launch of the {\it Fermi} satellite in mid-2008. In first few years it was remarkably stable with an integrated flux at $F\sim 2\E{-7}\,$ph/cm$^2$/s. In mid-2012 \cta\ exhibited a strong outburst with a peak flux of $F\sim 8\times 10^{-6}\,$ph/cm$^2$/s \citep{lea16}. Ever since the source has remained in active states without long returns to the old quiescence level. This behavior is also visible in long-term optical and X-ray light curves. Yet, all these events were dwarfed by an outburst lasting 4 months in late 2016 to early 2017. During the first half, fluxes in all bands rose steadily by at least one, in the optical case even more than 2 orders of magnitudes above previous states. \cta\ became one of the brightest \g-ray sources in the sky and was even visible by eye through small telescopes in the optical despite its redshift. Peak fluxes in the HE \g-ray band were at $F\sim 2\E{-5}\,$ph/cm$^2$/s. Over the next 2 months fluxes fell steadily back to pre-flare values resulting in an almost perfectly symmetric outburst. During the event, intra-night variability has been observed in both \g-rays \citep{sea18} and optical bands \citep{bea17,zea17}.

Optical, infrared and radio band data have been analyzed and interpreted by \cite{rea17} as an erratic wobbling of the jet resulting in different Doppler boosting of the emission. The required different boosting factors for optical, infrared and radio photons have been interpreted by the authors as signs for different locations of the emission regions of the respective wavelengths with optical close to the black hole and longer wavelengths progressively further down the jet. 
However, aside from not considering the $\gamma$-ray and X-ray lightcurves, the model does not provide an explanation for the required wobbling and large distances between individual emission regions.
% JYF: However, aside from not considering the $\gamma$-ray lightcurve, the model does not provide an explanation for the required wobbling and large distances between individual emission regions.
% MB: However, the model did not consider the gama-ray light curve, and it is unclear why regions of emission flaring simultaneously     should be quite far apart from each other and exhibit different erratic viewing angle fluctuations.
% AW: Also, the \g-ray light curve was not modeled. - why only \g-ray light curve is mentioned here?
% MZ: I used Felix' suggestion and added the X-rays.

In \cite{zea17}, hereafter paper I, we analyzed the possibility of the interaction of a gas cloud with the jet as the cause of \cta's flare. Such models were frequently used to explain fast flares \citep[e.g.,][]{bpb12,pmlh14,dcea17} on the order of hours or days by the interaction of the fully immersed object in the jet. In our framework, the material of the cloud is ablated by the ram pressure of the jet slice-by-slice as the cloud gradually enters the jet. The gas cloud cannot withstand the ablation process, since the jet's ram pressure is orders of magnitude stronger than the cloud's own gravitational pull. The ablation process leads to a gradual and symmetric change of the injection rate into the jet, which naturally and with minimal assumptions explained the long-term trend of \cta's flare. We modeled the spectra and light curves with a leptonic model using IC/BLR emission for the high-energy SED hump. 
% AW: In a previous paper -> maybe In first paper...
% MZ: Well, "first" could be interpreted as that there are more than 1 papers. I shortened it, now.

While the reproduction of the data was excellent, we were not able to conclusively identify the nature of the gas cloud. Derived parameters (density, size) did not match those of BLR clouds, which would have been a natural choice given that the emission region was located within the BLR. However, beyond the BLR the leptonic model would face some difficulty in explaining the high-energy component, since the radiation energy density of the DT emission is too low, and the above mentioned parameters can only be considered upper limits \citep{mea11}. In this paper we explore a hadronic model in order to account for the strong flare in \cta. Since the gas cloud does also provide protons, the injection is similar to the leptonic case. The hadronic model allows us to explore different locations of the emission region within the jet, which results in different parameters of the cloud. 

The outline of the paper is as follows. In section~\ref{sec:ana} we describe the data analysis, which is similar but updated from paper I. The detection of \g-ray photons with almost 100\,GeV by \fermi\ indicate that the intrinsic absorption by, e.g., the BLR cannot be severe and we derive a lower limit of the distance of the emission region from the black hole in section~\ref{sec:blr}. In section~\ref{sec:mod} we describe the detailed modeling. We reproduce the 4-months long outburst \change{considering three different distances from the black hole. We further attempt to reproduce six of the days-long flares on top of the longer trend, which is described in appendix \ref{sec:modIV}.} We discuss and interpret the results in section~\ref{sec:dis}, \change{considering the jet power, the neutrino output, the relation of our model to fast flares, and the inferred cloud parameters. Section \ref{sec:origin} is devoted to the nature of the gas cloud and how that might relate to the behavior of \cta\ during the last few years. We summarize our findings} in section~\ref{sec:sum}.

We use the following nomenclature: ``Long-term'' refers to the total 4-months long outburst. ``Medium-term'' refers to the days-long subflares on top of the long-term trend, while ``short-term'' refers to the fast flares on sub-hour time scales.
Primed quantities are in the AGN frame, quantities marked with the superscript ``obs'' are in the observer's frame, and unmarked quantities are in the comoving jet frame. We use a standard, flat cosmology with $H_0 = 69.6\,$km/s/Mpc, and $\Omega_M = 0.27$, which gives a luminosity distance $d_L = 2.19\times 10^{28}\,$cm.
%
%################################################################################################################################
%
\section{Data analysis} \label{sec:ana}
The data analysis in this paper has been extended compared to paper I in order to obtain more spectra. The procedures and updates are given below.
\begin{figure*}[th]
\centering 
\includegraphics[width=0.90\textwidth]{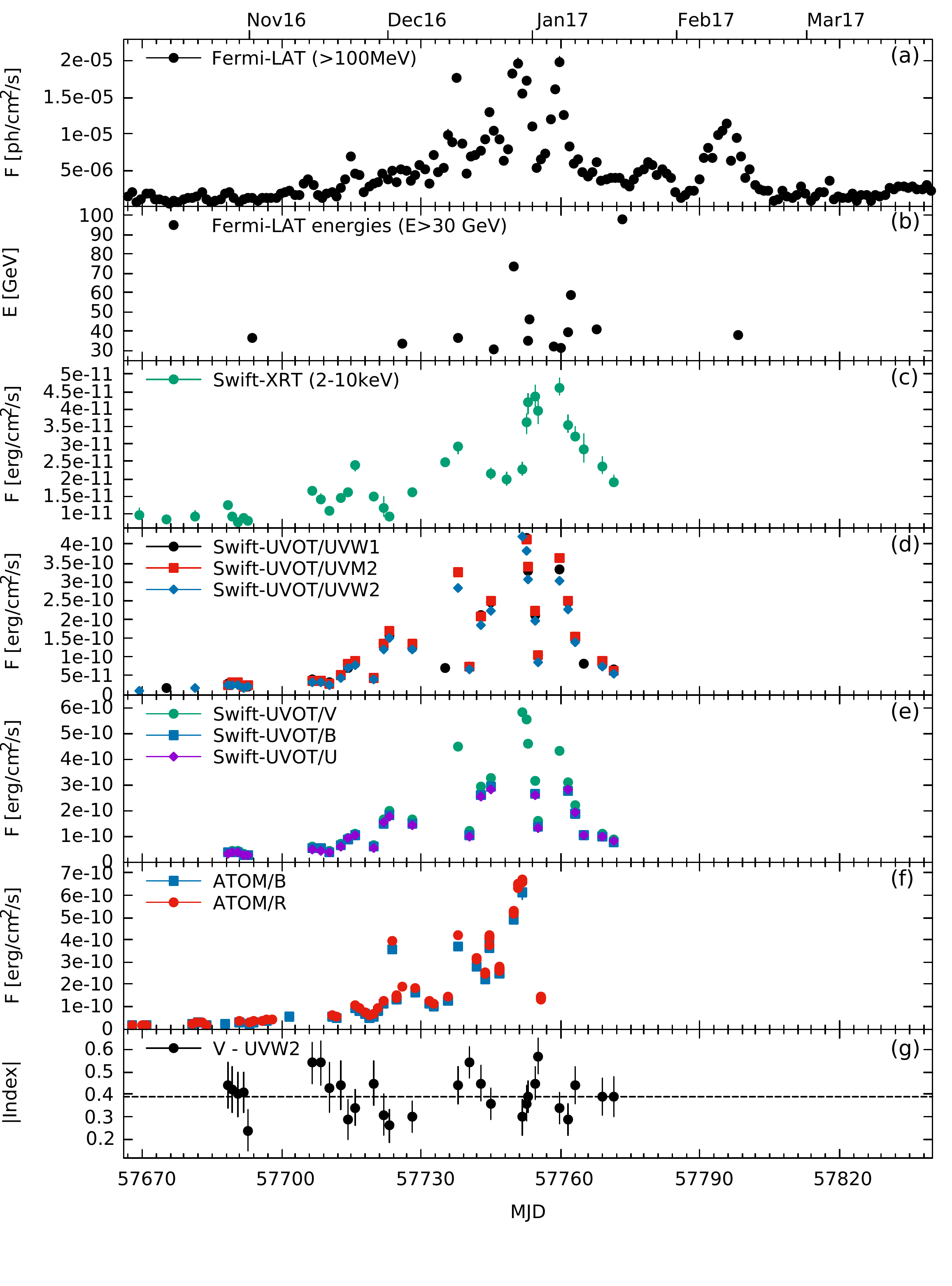}
\caption{As a function of time {\bf (a)} daily \fermi\ fluxes, {\bf (b)} photons detected with \fermi\ above 30\,GeV, {\bf (c)} {\it Swift}-XRT fluxes, {\bf (d)} {\it Swift}-UVOT UV fluxes, {\bf (e)} {\it Swift}-UVOT optical fluxes, and {\bf (f)} optical fluxes from ATOM as labeled. Panel {\bf (f)} gives the spectral index derived for {\it Swift}-UVOT data using V and UVW2 band fluxes with the dashed line marking the average. }
\label{fig:lc}
\end{figure*} 
%
% AW: Figure 1. - in the labels there should be LAT (not Fermi). Like it is in the case of XRT or UVOT.
% MZ: Thanks. I also found some inconsistencies in/to the other plots and corrected those.
%
\begin{figure*}[th]
\centering 
\includegraphics[width=0.98\textwidth]{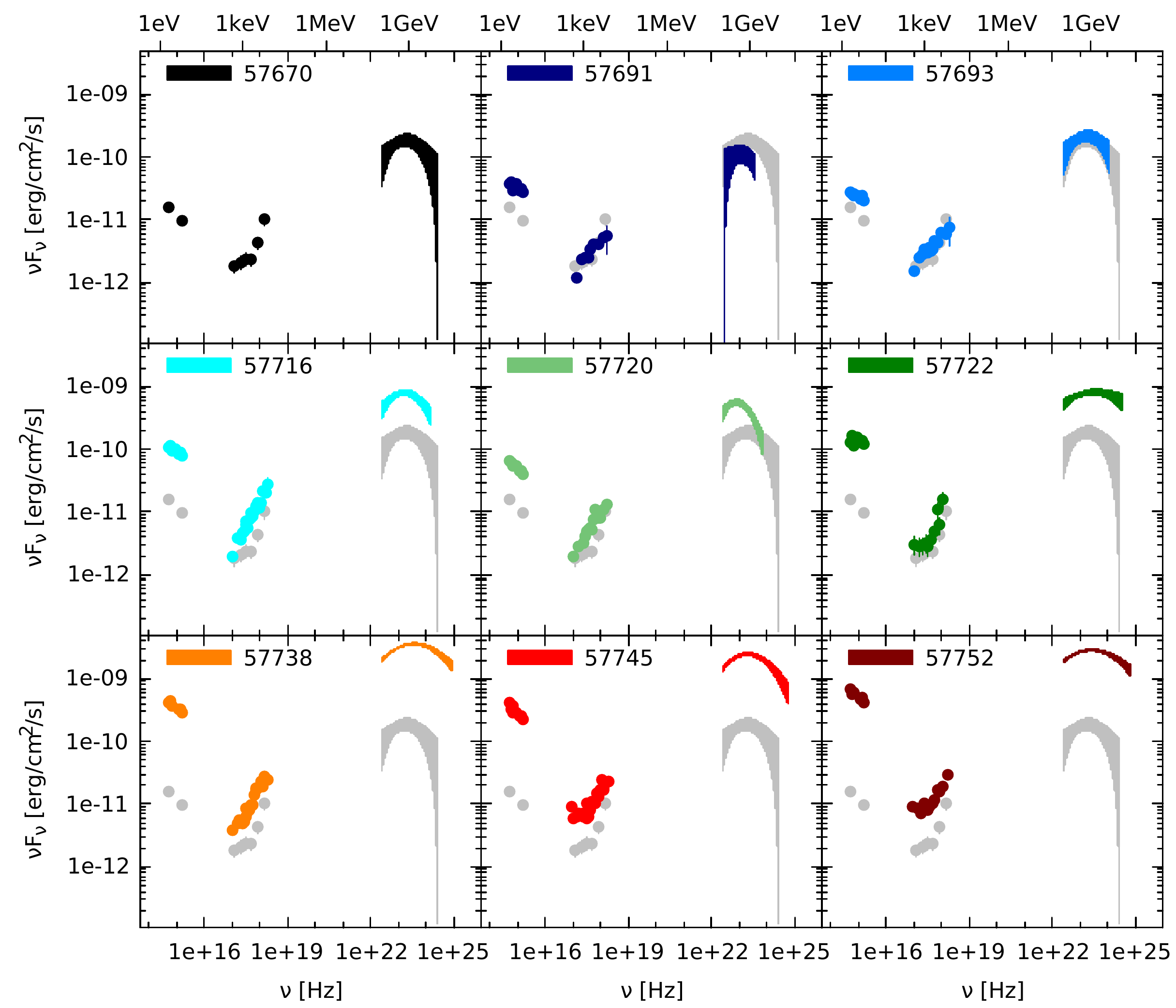}
\caption{Spectra for dates, where near-simultaneous data by all three observatories are available. The MJDs are indicated. The gray spectrum is the MJD 57670 spectrum for comparison.
}
\label{fig:spec-all}
\end{figure*} 
%
%% JPL: why are we using only the Fermi-LAT log-parabolic spectra, even if the LP does not significantly describe better the source than a PL ?
% MZ: The main reason is that we get a peak energy, which can be used to constrain model parameters. That one kind of expects the spectra to be curved is true, but probably even less convincing. 
% MB: Enlarge Fig 2
% MZ: Done. For both.

%###############################################################################################################################
\subsection{Fermi-LAT data analysis} \label{sec:fermi}
\begin{table}[th]
%\centering
\footnotesize
\caption{\textit{Fermi}-LAT detailed results of CTA\,102 used for each spectrum shown in Fig.~\ref{fig:spec-all}. The columns give the center of the daily-binned time interval in which the spectrum is extracted in MJD, the integrated flux density, the photon index, the curvature index, and the log-likelihood ratio between the two spectral hypothesis (log-parabola with respect to power-law).}
\begin{tabular}{ccccc}
MJD		& $F_\textnormal{100\,MeV--500\,GeV}$ & $\Gamma$ & $\beta$ & TS$_{\textnormal{LP}/\textnormal{PL}}$\\
        & $(\times 10^{-6} \textnormal{cm}^{-2} \textnormal{s}^{-1})$ & & & \\
\hline
57670  & $0.86 \pm 0.21$ & $1.673 \pm 0.339$ & $0.185 \pm 0.149$ & 2.6\\
57691  & $0.55 \pm 0.25$ & $1.763 \pm 0.596$ & $0.310 \pm 0.369$ & 1.7\\
57693  & $1.01 \pm 0.21$ & $1.689 \pm 0.289$ & $0.186 \pm 0.137$ & 2.7\\
57716  & $3.81 \pm 0.48$ & $1.732 \pm 0.162$ & $0.175 \pm 0.072$ & 6.4\\
57720  & $2.86 \pm 0.31$ & $1.976 \pm 0.151$ & $0.300 \pm 0.118$ & 9.5\\
57722  & $4.12 \pm 0.40$ & $1.811 \pm 0.119$ & $0.064 \pm 0.041$ & 2.9\\
57738  & $16.5 \pm 0.44$ & $1.738 \pm 0.036$ & $0.083 \pm 0.014$ & 46.8\\
57745  & $12.0 \pm 0.55$ & $1.758 \pm 0.062$ & $0.123 \pm 0.029$ & 20.7\\
57752  & $14.7 \pm 0.41$ & $1.821 \pm 0.037$ & $0.073 \pm 0.015$ & 1.6\\
\end{tabular}
\label{tab:fermi}
\end{table}
The LAT instrument \citep{aFea09} onboard the \textit{Fermi} satellite surveys the high energy \g-ray sky every 3 hours, with energies between 20\,MeV and above 300\,GeV, thus making it an ideal instrument to monitor the activity of \cta. This AGN has been reported in all the available \fermi\ catalogs, and is identified as \object{3FGL~J2232.5+1143} in the third \fermi\ source catalog \citep{aFea15}.

The \fermi\ data are analyzed using the public ScienceTools \texttt{v11r5p3}\footnote{See \href{http://fermi.gsfc.nasa.gov/ssc/data/analysis/documentation}{http://fermi.gsfc.nasa.gov/ssc/data/analysis/documentation}.}. Events in a circular region of interest of 10\degr\ in radius are extracted, centered on the nominal position of \object{3FGL~J2232.5+1143}. We will focus on high-energy spectra of the source at different epochs. Data are considered in the 100\,MeV--500\,GeV energy range. The \texttt{P8R2\_SOURCE\_V6} instrument response functions (event class 128 and event type 3) were used, together with a zenith angle cut of 90\degr\ to avoid contamination by the \g-ray bright Earth limb emission. The model of the region of interest was built based on the 3FGL catalog \citep{aFea15}. The Galactic diffuse emission has been modeled using the file \texttt{gll\_iem\_v06.fits} \citep{aFea16} and the isotropic background using \texttt{iso\_P8R2\_SOURCE\_V6\_v06.txt}. In the following, the source spectrum will be investigated both with a power-law shape

\begin{equation}
  \frac{dN}{dE} = N_0 \left(\frac{E}{E_0} \right)^{-\Gamma} \label{eq:pow} ,
\end{equation}
and a log-parabola
\begin{equation}
  \frac{dN}{dE} = N_0 \left(\frac{E}{E_0} \right)^{-(\Gamma + \beta \log(E/E_0))} \label{eq:logp},
\end{equation}
with $E_0=308$\,MeV fixed to the value reported in the 3FGL catalogue, the normalization $N_0$, photon index $\Gamma$, and spectral curvature $\beta$.

%% OLD TEXT: For the considered period between August 2015 and May 2017, \cta\ is detected with a Test Statistic \citep[TS,][]{mea96} of 163879, i.e. $\sim$405$\sigma$. The spectrum of \cta\ is significantly curved with a photon index of $\Gamma=2.068 \pm 0.008$ and a curvature index of $\beta=0.064 \pm 0.003$. The average flux is $F=(2.27 \pm 0.01) \times 10^{-6}$\,ph\,cm$^{-2}$\,s$^{-1}$.

For the main part of our work, we construct a daily-binned light curve, spanning from  October 1, 2016 to April 1, 2017. Since on daily time scales the preference of a log-parabola is not guaranteed, the spectrum has been modeled with a simple power-law in each time bin, leaving the photon index free to vary. The resulting light curve is shown in Fig.~\ref{fig:lc}(a).

From this data set, spectra were derived for different nights, chosen to be strictly simultaneous with \textit{Swift} and ATOM observations, presented in Fig.~\ref{fig:spec-all}. For each of these spectra, a log-parabolic shape is assumed, which parameters are reported in Table~\ref{tab:fermi}. In each case, the log-likelihood ratio with respect to a power-law hypothesis, TS$_{\textnormal{LP}/\textnormal{PL}}$, is also given. The maximum energy displayed for each spectrum (see also panel (b) in Fig.~\ref{fig:lc} and Tab.~\ref{tab:paramconst}) corresponds to the highest energy of photons attributed to CTA\,102 at more than 95\% CL, as evaluated using \texttt{gtsrcprob}. %The results for each spectrum are detailed in Table~\ref{tab:fermi}. 

\subsection{X-ray analysis} \label{sec:xray}
\begin{table}[th]
%\centering
\footnotesize
\caption{\textit{Swift}-XRT observations of CTA\,102 used for the spectra shown in Fig.~\ref{fig:spec-all}. The columns give the MJD, the Observation ID, the duration of the observation, as well as the spectral parameters, i.e. normalization $N_{0,\rm XRT}$ and index $\Gamma_{\rm XRT}$. The normalization energy is $E_0 = 1\,$keV.}
\begin{tabular}{ccccc}
MJD		& ObsID	& $t_{\rm dur}$ [ks] & $N_{0,\rm XRT}$ [ph/cm$^2$/s/keV]& $\Gamma_{\rm XRT} $ \\
\hline
57670	& 00033509084	& $0.6$	& $(1.2\pm0.2)\E{-3}$	& $1.3\pm0.2$ \\
57691 	& 00033509090	& $1.7$	& $(1.17\pm0.09)\E{-3}$	& $1.41\pm0.09$ \\
57693 	& 00033509092	& $6.0$	& $(1.56\pm0.09)\E{-3}$	& $1.55\pm0.07$ \\
57716	& 00033509098	& $12.1$	& $(2.6\pm0.1)\E{-3}$	& $1.20\pm0.05$ \\
57720 	& 00033509099	& $1.9$	& $(2.1\pm0.1)\E{-3}$	& $1.34\pm0.06$ \\
57722 	& 00033509100	& $0.4$	& $(1.9\pm0.3)\E{-3}$	& $1.4\pm0.2$ \\
57738	& 00033509106	& $2.4$	& $(3.2\pm0.1)\E{-3}$	& $1.2\pm0.5$ \\
57745 	& 00033509109	& $6.5$	& $(3.9\pm0.2)\E{-3}$	& $1.52\pm0.06$ \\
57752 	& 00033509111	& $1.8$	& $(5.1\pm0.2)\E{-3}$	& $1.63\pm0.06$ \\
\end{tabular}
\label{tab:XRT}
\end{table}
% AW: Table 2. - N_0 -> N_{0, XRT}
% MZ: Done.
%
The \change{Neil Gehrels} Swift Gamma-Ray Burst Mission  \citep[hereafter \textit{Swift},][]{gea04} is a multi-wavelength space instrument, which monitors sources in the optical, ultraviolet and X-ray energy bands. 
X-ray observations of \cta\ were possible thanks to  the X-ray Telescope \citep[XRT,][]{bea05} onboard. It monitored \cta\ since 2005 in 144 pointing observations taken in the energy range of 0.3-10\,keV.
For the goal of this paper, data collected between MJD 57668 and MJD 57821, which correspond to the ObsIDs of 00033509084-00033509120 have been analyzed and are presented in the light curve (Fig.~\ref{fig:lc}(c)).

X-ray data analysis was performed using version 6.20 of the HEASOFT package.\footnote{\url{http://heasarc.gsfc.nasa.gov/docs/software/lheasoft}}
The data were recalibrated using the standard procedure \verb|xrtpipeline|.
Spectral fitting was carried out using \verb|XSPEC| v.12.8.2 \citep{a96}.
Data were binned in order to have at least 30 counts per bin and were fitted using a power-law model, Eq. (\ref{eq:pow}), with the Galactic absorption value of $N_{H} = 4.76 \times 10^{20}$\,cm$^{-2}$ \citep{kea05} set as a frozen parameter. 
Furthermore, it was tested whether the spectrum can be better descibed with a broken power-law  model.
According to reduced $\chi^2$ values, a simple power-law is preferred for all X-ray spectra of \cta. 

The observations presented in the SEDs (Fig.~\ref{fig:spec-all}) are summarized in Tab.~\ref{tab:XRT}. While there is some variability in the spectral index, this is mostly minor and most of the X-ray variability comes from a change in the normalization.

%###############################################################################################################################
\subsection{Optical/UV analysis} \label{sec:optical}
In the optical and ultraviolet regime, \cta\ was monitored with the UVOT instrument onboard \textit{Swift}.
The monitoring was performed in the bands UVW2 (188 nm), UVM2 (217 nm), UVW1 (251 nm), U (345 nm), B (439 nm), and V (544 nm), with the number in brackets giving the central wavelengths.
The \verb|uvotsource| task was used to calculate the instrumental magnitudes. In this case  all photons from a circular region with radius 5'' were taken into account. 
The background was determined from the same size region, located close to \cta\ and not being contaminated with signal from any nearby source.
Data were corrected for dust absorption using the reddening $E(B-V)$ = 0.0612\,mag \citep{sf11} and the ratios of the extinction to reddening, $A_{\lambda} / E(B-V)$ \citep{gea06}. 
The results of the UVOT monitoring are shown in Fig.~\ref{fig:lc}(d) and (e).

Further optical data in R- and B-band filters have been obtained with the Automatic Telescope for Optical Monitoring (ATOM), which is a $75\,$cm optical telescope located at the H.E.S.S. site in the Khomas Highland in Namibia \citep{hea04}. 
%It regularly observes roughly 300 $\gamma$-ray emitters.

ATOM monitors \cta\ since 2008. 
During the visibility period presented in this paper, R-band monitoring lasted from June 2016 until January 2017. 
Additional B-band observations were taken from October 2016 until December 2016.
Most of the high-flux period is covered by at least one B-band and several R-band measurements per night.
The data were analyzed using the fully automated ATOM Data Reduction and Analysis Software and have been manually quality checked.
The resulting flux was calculated via differential photometry using 5 custom-calibrated secondary standard stars in the same field-of-view.
%
%Using measurements from a calm period between 2008 and 2011 the baseline flux of \cta\ can be established as $R = 16.90 \pm 0.02\,$mag.
%An outburst in September 2012 reached $R = 14.6 \pm 0.1$mag before returning to previous levels. In late 2015, ATOM detected \cta\ at $R = 16.54 \pm 0.08\,$mag.
%
%Beginning in mid 2016, \cta\ showed increasing activity with a first outburst in August reaching $R = 14.20 \pm 0.02\,$mag.
%Towards the end of visibility \cta\ started to steadily brighten, culminating in $R = 10.96 \pm 0.05\,$mag on 29 December 2016 (MJD 57751). We find significant intra-night variability, similar to the results reported in \cite{bea17}.
Both R- and B-band light curves are shown in Fig.~\ref{fig:lc}(f). %The fastest optical variability was detected in the R-band on MJD~57747 with a variability time (following the definition of \cite{zea99}) of $0.03\,$d$\sim 45\,$min.

We have used the V and UVW2 fluxes from {\it Swift}-UVOT to derive the spectral index of the optical-UV SED. This can be calculated from
\begin{align}
\alpha = \frac{\log{\left( \nu F_{\nu,\mbox{UVW2}} \right)} - \log{\left( \nu F_{\nu,\mbox{V}} \right)}}{\log{\left( \nu_{\rm UVW2} \right)} - \log{\left( \nu_{\rm V} \right)}} \label{eq:optind},
\end{align}
where $\nu F_{\nu,i}$ are the SED fluxes in the respective bands and $\nu_i$ are the central frequencies of the filters. The resulting indices are plotted in Fig.~\ref{fig:lc}(g). While there are some variation around the average of $\sim 0.4$, these are not statistically significant ($p$-value of a constant $\sim 40\%$). For the modeling, we assume a constant optical-UV SED index.

%###############################################################################################################################
\subsection{Analysis of long-term data} \label{sec:ltdata}
Additionally to the flare data, we have also analyzed the long-term data of \cta. This includes all data taken with \fermi\, {\it Swift}-XRT, and ATOM spanning from  August 5, 2008 to \change{November 01}, 2018. %% UPDATE WITH FINAL DATE
The same analysis steps as detailed above were followed to create the light curves. The results are used for discussion in \change{section \ref{sec:origin}.} 

Note that \cta\ is not observable for {\it Swift} and ATOM between mid-January and late-April each year. Furthermore, it is not a regular target for {\it Swift} resulting in large gaps between observations. 
% To Felix & Stefan: Mention ATOM technical problems in 2013 and 2014?
% JYF: No. It's already visible in figure 7.

%
%###############################################################################################################################
\section{Absorption by the BLR} \label{sec:blr}
\begin{figure*}[th]
\begin{minipage}{0.49\linewidth}
\centering \resizebox{\hsize}{!}
{\includegraphics{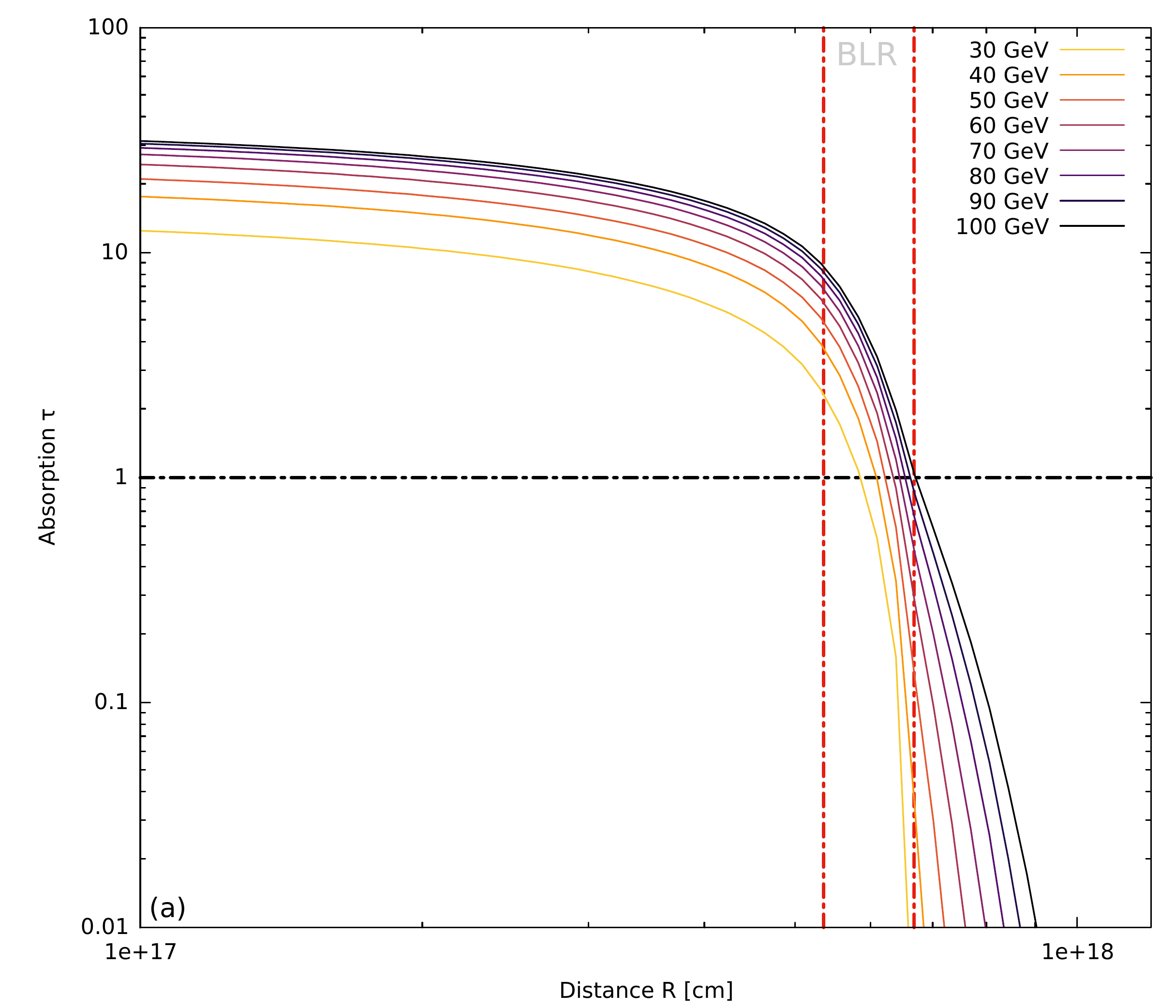}}
\end{minipage}
\hspace{\fill}
\begin{minipage}{0.49\linewidth}
\centering \resizebox{\hsize}{!}
{\includegraphics{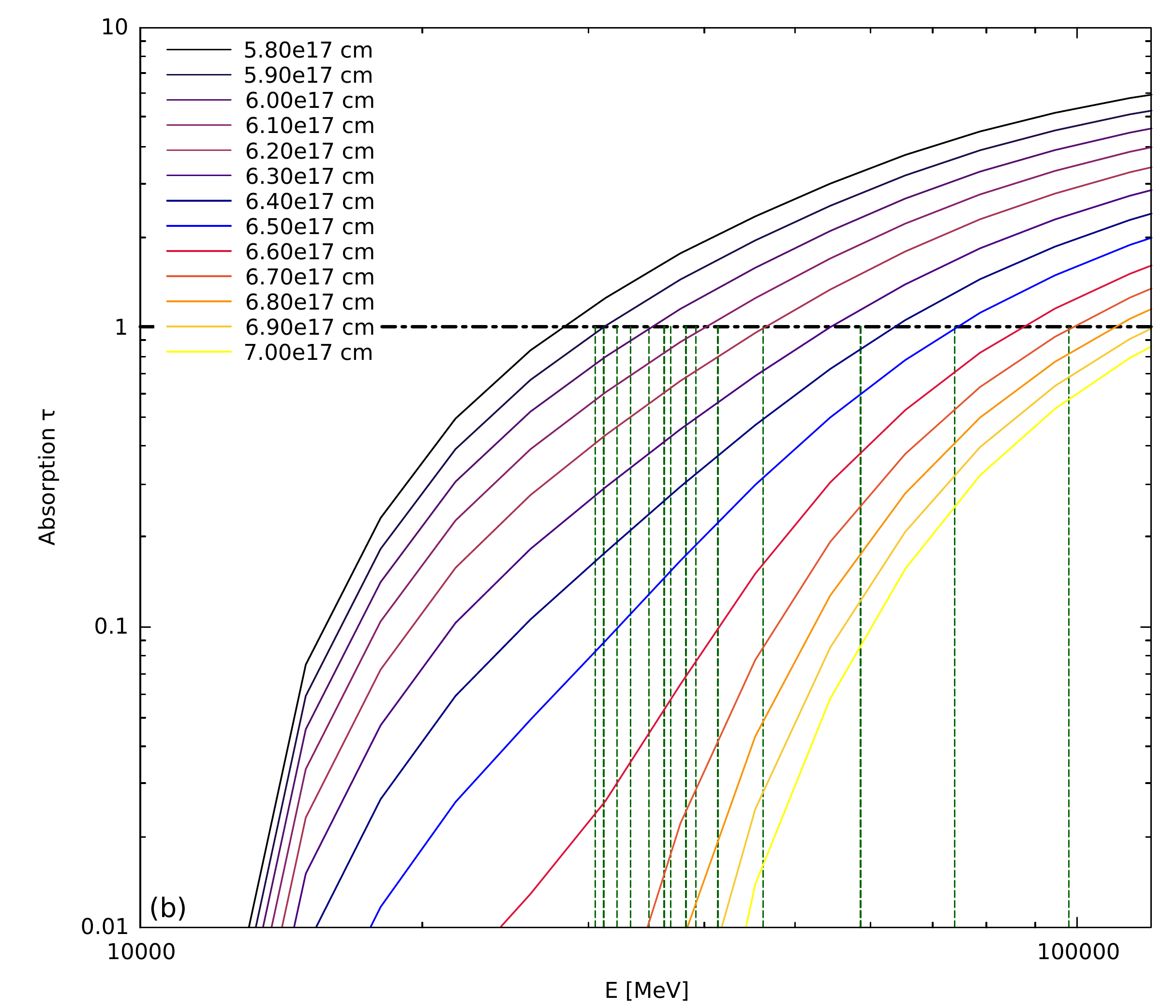}}
\end{minipage}
\caption{
{\bf (a)} Absorption $\tau$ through the BLR as a function of distance for observed $\gamma$-rays with energies as labeled. The vertical red dot-dashed lines mark the inner and outer boundary of the BLR, while the horizontal black dot-dashed line marks $\tau=1$.
{\bf (b)} Absorption $\tau$ through the BLR as a function of observed energy for different distances as labeled. The vertical green dashed lines mark photon energies detected with \fermi\ with $E>30\,$GeV, while the horizontal black dot-dashed line marks $\tau=1$.
}
\label{fig:blrabs}
\end{figure*} 
In paper I we used the IC/BLR process in order to reproduce the high-energy component. However, this process requires the emission region to be located within the BLR, and the BLR photons, while good targets for inverse-Compton scattering, are also good absorbers of $\gamma$-rays through $\gamma\gamma$ pair production. In order to keep the absorption small, we placed the emission region close to the outer edge of the BLR in paper I.

Here, the strength of the absorption by the BLR for different energies is investigated. By comparing it to the energies of photons with $E>30\,$GeV, c.f. Fig.~\ref{fig:lc}(b), we can place constraints on the distance of the emission region from the black hole. In order to evaluate the absorption from the BLR with the parameters listed in the introduction, we use the code by \cite{be16} that models the BLR as a thin shell between $0.8\times R_{\rm BLR}$ and $R_{\rm BLR}$ and calculates the absorption by considering all soft photon paths through the BLR depending on the location of the $\gamma$-ray photon and the incident angle. The absorption value $\tau$ is plotted as a function of distance in Fig. \ref{fig:blrabs}(a), and as a function of energy in Fig. \ref{fig:blrabs}(b). Within the shell, photons with energy $\gtrsim 30\,$GeV are absorbed, while the absorption is quickly reduced once the emission region crosses through the BLR shell. The absorption through the DT photon field is negligible at energies below $100\,$GeV.

With the detection of $\gamma$-ray photons with $E>30\,$GeV, the emission region cannot be deep within the BLR, proving a posteriori our assumption in paper I. This is also in line with findings by \cite{cea18}. Whether the individual and rather singular detections of photons with $E>50\,$GeV imply that the emission region must be outside of the BLR, is difficult to say. A clear answer would have only been possible with a larger number of photons, since single photons might always be the lucky ones that escape absorption. The effective area of \fermi\ is too small to provide a satisfactory answer, and we will have to wait for the future Cherenkov Telescope Array \citep{acha13} to explore this question in greater detail for \cta.

These considerations ignore the possibility of a pair cascade initiated by $\gamma\gamma$ absorption within the BLR \citep[e.g.,][and references therein]{rb12}. If the emission region would be located within the BLR, $\gamma\gamma$ absorption would lead to strong attenuation of the \g-ray flux above a few GeV, which would be re-emitted in the sub-GeV regime through the cascade. In this case, the resulting MeV -- GeV \g-ray spectrum would be much softer than the observed Fermi-LAT spectrum. As we have placed the \g-ray emission region near the outer edge of the BLR or beyond, $\gamma\gamma$ absorption affects at most a small fraction of the $\gtrsim 10$~GeV flux, whose re-emission at lower energies through pair cascades will make a negligible contribution to the total spectrum.
%Nevertheless, we can conclude that the emission region is at least on the edge of the BLR or beyond. In fact, being located (far) beyond the BLR would mean that the cloud that gets devoured by the jet, would be moving slower. In turn, the size of the cloud would be smaller and the temperature would rise -- probably making the cloud more realistic in terms of the latter parameter. However, electrons would quickly run out of target photons for the IC process. While the dusty torus is an option, it is usually not considered a good option when it comes to fast flares, since the cooling is much slower than with BLR photons. In order to be independent on the specific target photons, we try a hadronic model, where the external photon fields play a reduced (yet not negligible) role.

%
%###############################################################################################################################
\section{Modeling} \label{sec:mod}
In order to reproduce the long-term evolution of the outburst in \cta, we use the same cloud ablation model as in paper I. A brief summary of the process that leads to the overall injection form is presented in appendix \ref{app:theory}.
The hadronic model is based upon the code developed by \cite{dbf15}. It is a one-zone code that calculates self-consistently the particle distributions and the emerging photon spectra. It does not consider the pre-acceleration of particles, which might happen in a small acceleration zone as in the models of \cite{ws15} and \cite{cpb15}. We have added new features that enhance the possibilities of the \cite{dbf15} code, namely the inclusion of external photon fields for absorption as in \cite{be16} and as target fields for protons (pion production) and electrons (inverse Compton process). Details of the code, its parameters and our additions are described in appendix \ref{app:hadmod}. There we also show a plot with a detailed spectrum explaining how the total spectra are created from the individual processes.

%###############################################################################################################################
\subsection{Constraints and generic parameters} \label{sec:cons}
%
%List of constraints:
% - Spectral indizes (protons, electrons) !!!DONE!!!
% - Variability time scale !!!DONE!!!
% - Break between X-rays and X-ray to gamma-ray domain (-> B*gamma_min^2) ???
% - Peak energy in gamma-rays (-> B*gamma_max^2) !!!DONE!!!
%
\begin{table*}[th]
\caption{Observational constraints on parameters. [1] MJD of the derived SED; [2] The X-ray spectral index; [3] The spectral index between the Swift-XRT and \fermi\ SED; [4] The proton spectral index; [5] The maximum energy of the \fermi\ SED in GeV; [6] The numerical value of $F_{\rm B\gamma}$ in Gauss; \\ A pair of [7] the magnetic field and [8] the maximum proton Lorentz factor that fulfills $F_{\rm B\gamma}$.}
\begin{tabular}{c|ccc|cccc}
MJD	& $\alpha_{\rm X}$	& $\alpha_{\rm MeV}$	& $s_{\rm p}$	& $E_{\rm max,HE}$ [GeV]	& $F_{\rm B\gamma}$ [G]	& $B$ [G]	& $\gamma_{\rm p,max}$ \\
$[1]$	& $[2]$	& $[3]$	& $[4]$	& $[5]$	& $[6]$	& $[7]$	& $[8]$ \\
\hline
57670	& $0.8 \pm 0.2$	& $0.22 \pm 0.07$	& $2.55 \pm 0.14$	& $0.7$	& $6.0\times 10^{19}$	& $60$	& $1.0\times 10^{9}$ \\
57691	& $0.59 \pm 0.085$	& $0.25 \pm 0.13$	& $2.51 \pm 0.26$	& $0.5$	& $3.6\times 10^{19}$	& $50$	& $8.5\times 10^{8}$ \\
57693	& $0.45 \pm 0.07$	& $0.29 \pm 0.07$	& $2.43 \pm 0.15$	& $0.7$	& $5.7\times 10^{19}$	& $60$	& $9.7\times 10^{8}$ \\
57716	& $0.8 \pm 0.05$	& $0.29 \pm 0.04$	& $2.42 \pm 0.09$	& $0.7$	& $5.3\times 10^{19}$	& $60$	& $9.4\times 10^{8}$ \\
57720	& $0.66 \pm 0.06$	& $0.35 \pm 0.03$	& $2.30 \pm 0.07$	& $0.3$	& $2.6\times 10^{19}$	& $50$	& $7.2\times 10^{8}$ \\
57722	& $0.6 \pm 0.2$	& $0.35 \pm 0.04$	& $2.29 \pm 0.07$	& $1.4$	& $1.1\times 10^{20}$	& $70$	& $1.3\times 10^{9}$ \\
57738	& $0.8 \pm 0.5$	& $0.46 \pm 0.02$	& $2.08 \pm 0.05$	& $1.5$	& $1.2\times 10^{20}$	& $80$	& $1.2\times 10^{9}$ \\
57745	& $0.48 \pm 0.06$	& $0.44 \pm 0.02$	& $2.13 \pm 0.04$	& $0.8$	& $6.6\times 10^{19}$	& $70$	& $9.7\times 10^{8}$ \\
57752	& $0.37 \pm 0.06$	& $0.44 \pm 0.02$	& $2.13 \pm 0.03$	& $1.1$	& $8.5\times 10^{19}$	& $70$	& $1.1\times 10^{9}$ 
\end{tabular}
\label{tab:paramconst}
\end{table*}
\begin{table}[th]
\caption{Generic and initial parameters used in all models. The parameters below the horizontal line describe the external fields.}
\begin{tabular}{lcc}
Definition									& Symbol 					& Value \\
\hline
Doppler factor								& $\delta$					& $35\,$ \\ 
Emission region radius						& $R$						& $2.0\times 10^{16}\,$cm \\ 
Initial Magnetic field						& $B$						& $60\,$G \\ 
Minimum proton Lorentz factor				& $\gamma_{\rm p,min}$		& $1.0\times 10^6\,$ \\ 
Initial maximum proton Lorentz factor		& $\gamma_{\rm p,max}$		& $1.0\times 10^9\,$ \\ 
Proton spectral index						& $s_{\rm p}$				& $2.4\,$ \\ 
Minimum electron Lorentz factor				& $\gamma_{\rm e,min}$		& $2.0\times 10^2\,$ \\ 
Maximum electron Lorentz factor				& $\gamma_{\rm e,max}$		& $3.0\times 10^3\,$ \\ 
Electron spectral index						& $s_{\rm e}$				& $2.8\,$ \\ 
Escape time scaling							& $\eta_{\rm esc}$			& $5.0\,$ \\ 
Acceleration to escape time ratio			& $\eta_{\rm acc}$			& $30.0\,$ \\ 
\hline
Accretion disk luminosity					& $L_{\rm AD}\p$			& $3.75\times 10^{46}\,$erg/s \\ 
Radius of the BLR							& $R_{\rm BLR}\p$			& $6.7\times 10^{17}\,$cm \\
Temperature of the BLR						& $T_{\rm BLR}\p$			& $1.0\times 10^4\,$K \\
Luminosity of the BLR						& $L_{\rm BLR}\p$			& $4.14\times 10^{45}\,$erg/s \\
Radius of the DT							& $R_{\rm DT}\p$			& $6.18\times 10^{18}\,$cm \\
Temperature of the DT						& $T_{\rm DT}\p$			& $1.2\times 10^3\,$K \\
Luminosity of the DT						& $L_{\rm DT}\p$			& $7.0\times 10^{45}\,$erg/s 
\end{tabular}
\label{tab:genericparam}
\end{table}
From the observations described in section \ref{sec:ana}, we can derive constraints on the parameters of the emission region. These parameters are the spectral indices of the proton and electron distributions, the size of the emission region, the magnetic field and the maximum proton Lorentz factor.

The spectral index of the proton distribution\footnote{The spectral index $s$ of a particle distribution $n$ is defined as $n(\gamma)\propto \gamma^{-s}$, where $\gamma$ is the particle Lorentz factor.} $s_{\rm p}$ can be derived from the high-energy SED component by assuming that a simple power-law of the form $\nu F_{\nu}\propto \nu^{\alpha}$ connects the {\it Swift}-XRT SED and the \fermi\ SED. The resulting indices $\alpha_{\rm MeV}$ (the subscripts indicates that the index covers mostly the MeV domain of the SED) are listed in column 3 of Tab. \ref{tab:paramconst}. Assuming further that the proton cooling at Lorentz factors $\gamma$ corresponding to these photon energies is in the slow regime, which we verify a posteriori, the SED index and the proton spectral index are related by $s_{\rm p} = 3-2\alpha_{\rm MeV}$. The resulting proton indices are listed in column 4 of Tab. \ref{tab:paramconst}. The proton distribution hardens significantly from $s_{\rm p}\sim 2.4$ at the beginning of the flare to $s_{\rm p}\sim 2.1$ at the peak of the flare.

A similar strategy is used to obtain the electron spectral index $s_{\rm e}$ through the index of the optical-UV SED. As indicated by Fig.~\ref{fig:lc}(g), there is no significant variation in this parameter. Using the average optical-UV SED index of $\alpha_{\rm opt}\sim -0.4$, and assuming that electrons cool fast -- which is verified a posteriori, again -- the electron spectral index becomes $s_{\rm e} = 2-2\alpha_{\rm opt} = 2.8$.

The size of the emission region can be estimated through the observed minimum variability time scale. The latter, however, depends strongly on the cadence of observations, and the definition of the variability time scale. Here, we follow the definition of \cite{zea99}, which compares two subsequent flux points to estimate the variability time:
\begin{align}
t_{\rm var} = \frac{F_i+F_{i+1}}{2} \frac{t_{i+1}-t_{i}}{|F_{i+1}-F_{i}|} . \label{eq:tvardef}
\end{align}
The $\gamma$-ray light curve is binned in 1\,d intervals, which allows us to detect a minimum variability time scale of $0.93\,$d. Analyzing a couple of the brightest $\gamma$-ray flares in much greater detailed, allowed \cite{sea18} to detect variability on the order of a few minutes. Since we aim for an explanation of the long-term trend, we disregard this-short term variability. During the brightest subflares in late December 2016 / early January 2017, the observation cadence of Swift had been increased substantially to up to a few pointings per day. These observations reveal a minimum variability time scale in the X-ray domain of $1.62\,$d, while the fastest variability revealed by UVOT observations is $0.88\,$d. The cadence of ATOM observations had also been increased a lot during the brightest state of the source. These reveal a very fast variability in the R-band of $0.03\,$d in line with \cite{bea17}. B-band cadence was not as high as the R-band cadence and reveals a variability time scale of $0.48\,$d. As mentioned before, we focus on the long-term trend, and not on short time scales.\footnote{\change{Nonetheless, a discussion on the fast flares and how they fit into our general picture is provided in section \ref{sec:fastflares}.}} However, in order to incorporate at least the possibility to account for the medium time scales of the subflares within a single framework, we chose the half-day variability time scale as a representative of the emission region, giving
\begin{align}
 R = 2.0\times 10^{16} \est{\delta}{35}{}\ \mbox{cm}, \label{eq:tvar}
\end{align}
% MB: Equ. (4): Should have been R_b (not t_{var}).
% MZ: Done.
where $\delta$ is the Doppler factor of the emission region. This assumes that the emission region keeps the same size over the entire event, which is a standard assumption for the one-zone model. Note that the reproduction of the long-term trend is a consequence of our injection scenario and not of the size of the emission region. This is different for the medium-term subflares, where cooling and escape time scales play a more prominent role.

The energy of the peak of the high energy component is directly related to the underlying magnetic field and the maximum proton Lorentz factor: $E_{\rm max,HE}\propto B\gamma_{\rm p,max}^2$. Assuming a constant Doppler factor, we can estimate how this relation changes by measuring the energies of the maximum flux in the \fermi\ SEDs. Setting $F_{\rm B\gamma,i} = B_i \gamma_{\rm p,max,i}^2$, where the index $i$ is the MJD of the spectrum, we can derive this product for all dates as
\begin{align}
F_{\rm B\gamma,i} = F_{\rm B\gamma,57560} \frac{E_{\rm max,HE,i}}{E_{\rm max,HE,57560}} \label{eq:FBg}.
\end{align}
The energies are listed in column 5 of Tab.~\ref{tab:paramconst}. Assuming for the beginning of the flare $B_{\rm 57560}=60\,$G and $\gamma_{\rm p,max,57560}=1.0\times 10^9$, the resulting $F_{\rm B\gamma,i}$ are listed in column 6 of the same table, while the columns 7 and 8 give potential realizations for $B_i$ and $\gamma_{\rm p,max,i}$.

Following a similar strategy, one could in principle constrain the minimum proton Lorentz factor, as well. However, the X-ray spectra from Swift do not show a break, which could pinpoint the corresponding energy. On the other hand, the spectral indices of the X-ray spectra ($\alpha_{\rm X}$) and the indices from the extrapolated spectra in the MeV domain ($\alpha_{\rm MeV}$), as listed in columns 2 and 3 of Tab.~\ref{tab:paramconst}, indicate that there must be a break somewhere beyond $10\,$keV. This corresponds roughly to a minimum proton Lorentz factor of $\gamma_{\rm p,min}\sim 10^6$.

For the parameters of the electron distribution we can only derive upper limits except for the spectral index. The maximum electron Lorentz factor cannot be too large, or else the electron synchrotron component would influence the X-ray spectrum, which is not observed. Assuming a magnetic field of $60\,$G and a Doppler factor of $\delta=35$, the maximum and minimum electron Lorentz factors must be smaller than $4\times 10^3$ and $2.6\times 10^2$, respectively. \change{The latter is derived from the condition that the synchrotron peak is located below the R-band.}

There are a few more free parameters that are set (initially) for each model. The Doppler factor $\delta$ is set to $35$. This value is within the allowed range of observed superluminal motions of up to $18c$ observed in the \cta\ jet \citep{leaM16}. The magnetic field $B$ is set to $60\,$G to account for the necessary magnetic field strength to confine the protons to the emission region. The escape time scale is set as a multiple $\eta_{\rm esc} = 5.0$ of the light-travel time. The acceleration time is assumed as an energy-independent multiple $\eta_{\rm acc}=30.0$ of the escape time scale. The parameters of the external photon fields have been described in the introduction. All generic and/or initial parameters are listed in Tab.~\ref{tab:genericparam}.

%###############################################################################################################################
\subsection{Results} \label{sec:res}
%
% - How many models to show? Different distances (3), with all flares (at least 1)?
% - Discuss particle and cooling behavior
% - Discuss powers!
%
\begin{figure*}[th]
\centering 
\includegraphics[width=0.95\textwidth]{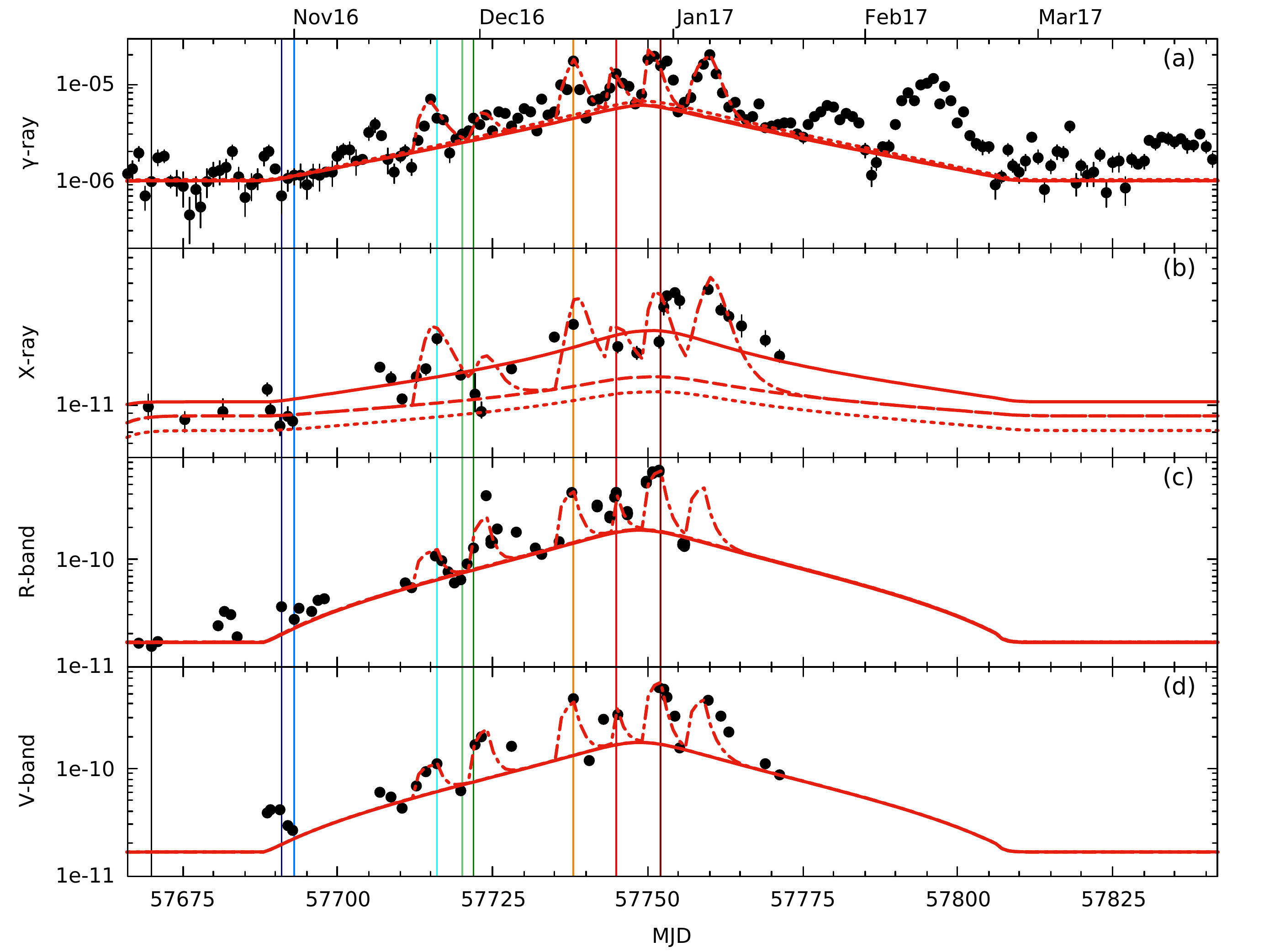}
\caption{Light curves of {\bf (a)} \fermi\ data, {\bf (b)} {\it Swift}-XRT data, {\bf (c)} ATOM/R, and {\bf (d)} {\it Swift}-UVOT/V data. The thick red lines are the modeling result for Examples I (solid), II (dashed), and III (dotted). \change{The dot-dashed lines mark the Example IV, for which details are provided in appendix \ref{sec:modIV}.} The vertical thin lines mark the dates, where the spectra have been extracted (same color code as in Fig.~\ref{fig:spec-all}). Note the logarithmic scaling of the y axis.}
\label{fig:mod-lc}
\end{figure*} 
\begin{figure*}[th]
\centering
\includegraphics[width=0.95\textwidth]{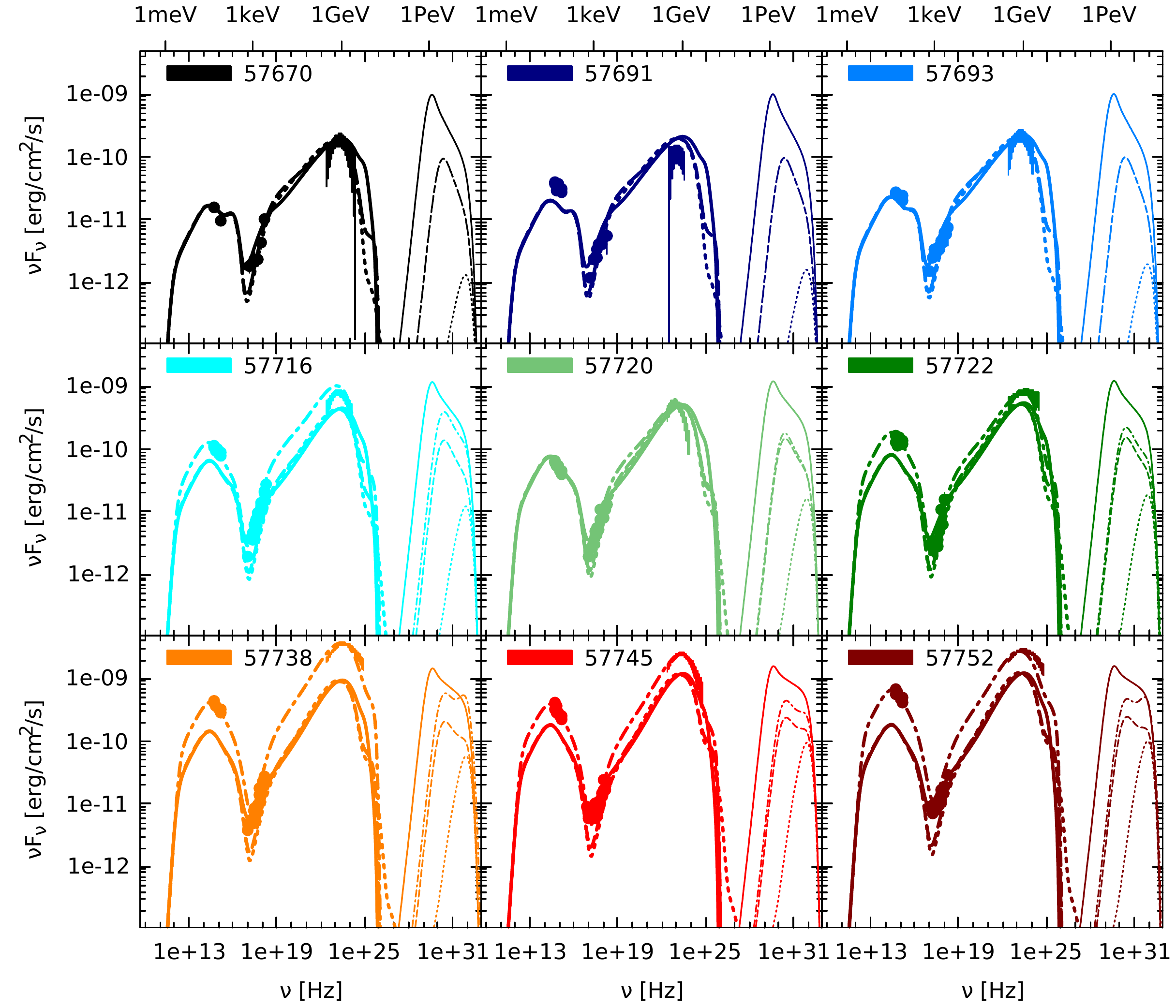}
\caption{SED and models for the dates with data by all instruments. The $\gamma$-ray data has been corrected for EBL-absorption following the model by \cite{frv08}. The thick lines mark the photon spectra, while the thin lines mark the neutrino spectra. \change{Line styles are the same as in Fig.~\ref{fig:mod-lc}.}}
\label{fig:mod-spec}
\end{figure*} 
\begin{figure*}[th]
\begin{minipage}{0.49\linewidth}
\centering \resizebox{\hsize}{!}
{\includegraphics{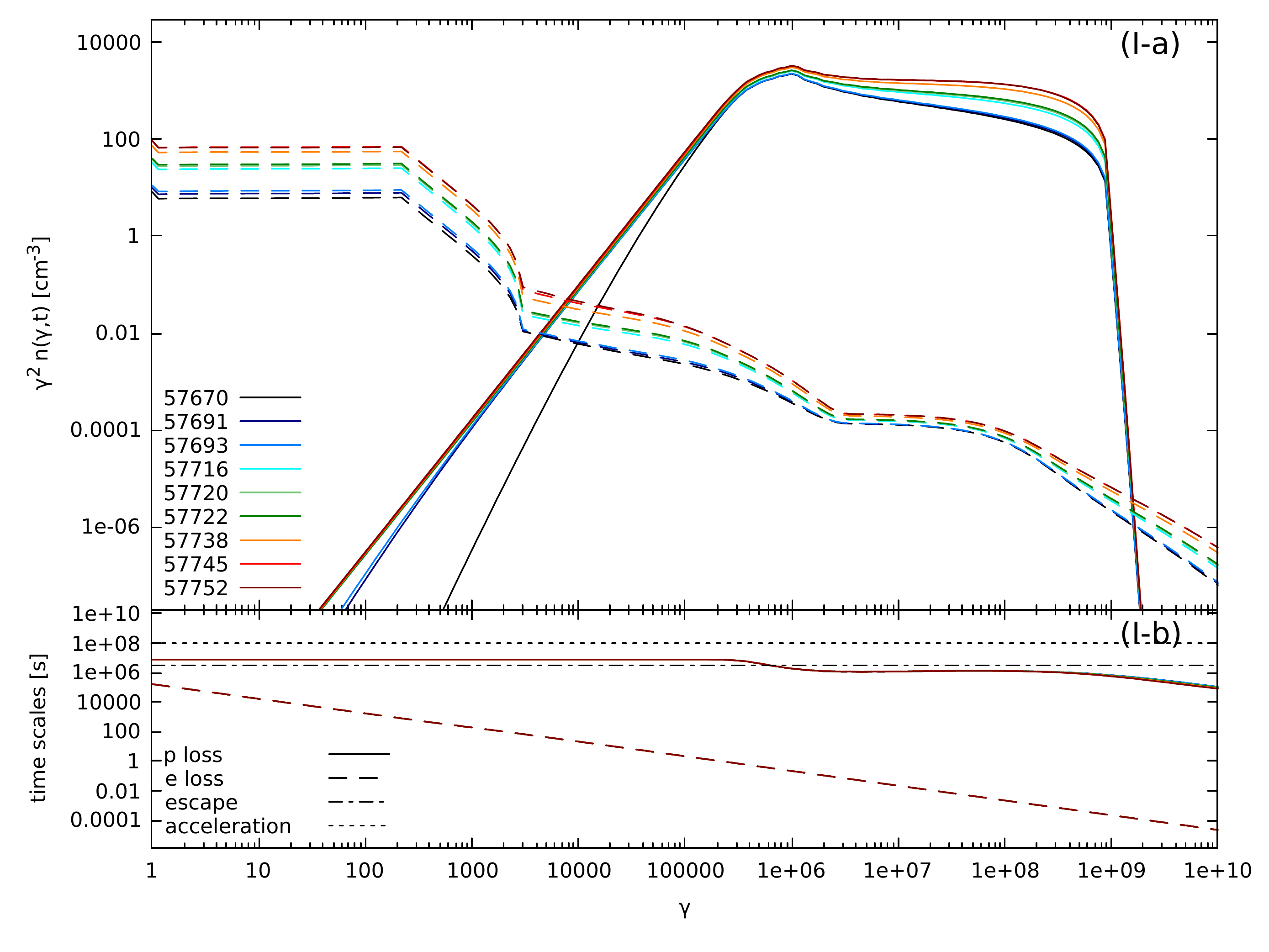}}
\end{minipage}
\hspace{\fill}
\begin{minipage}{0.49\linewidth}
\centering \resizebox{\hsize}{!}
{\includegraphics{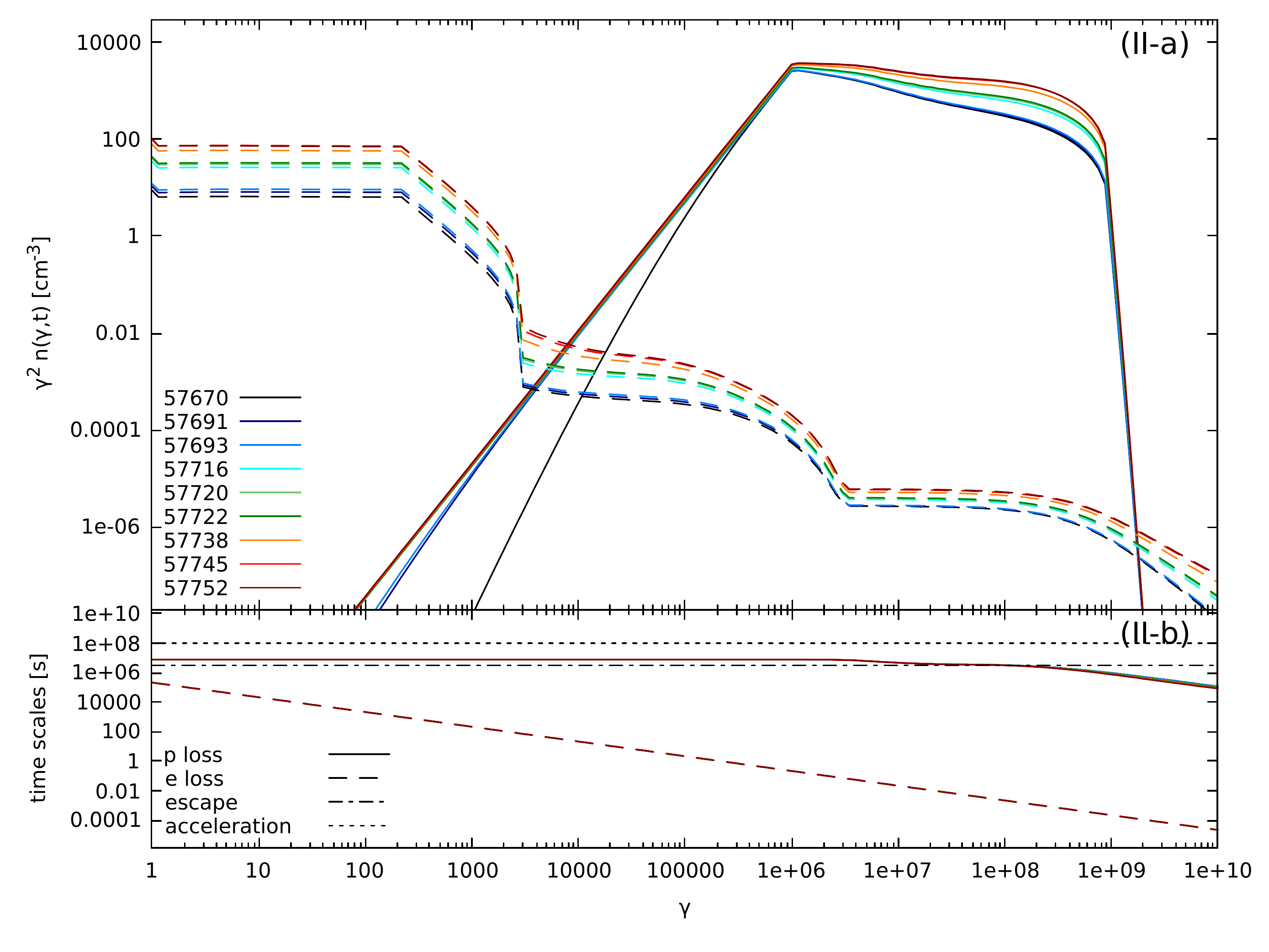}}
\end{minipage}
\newline
\begin{minipage}{0.49\linewidth}
\centering \resizebox{\hsize}{!}
{\includegraphics{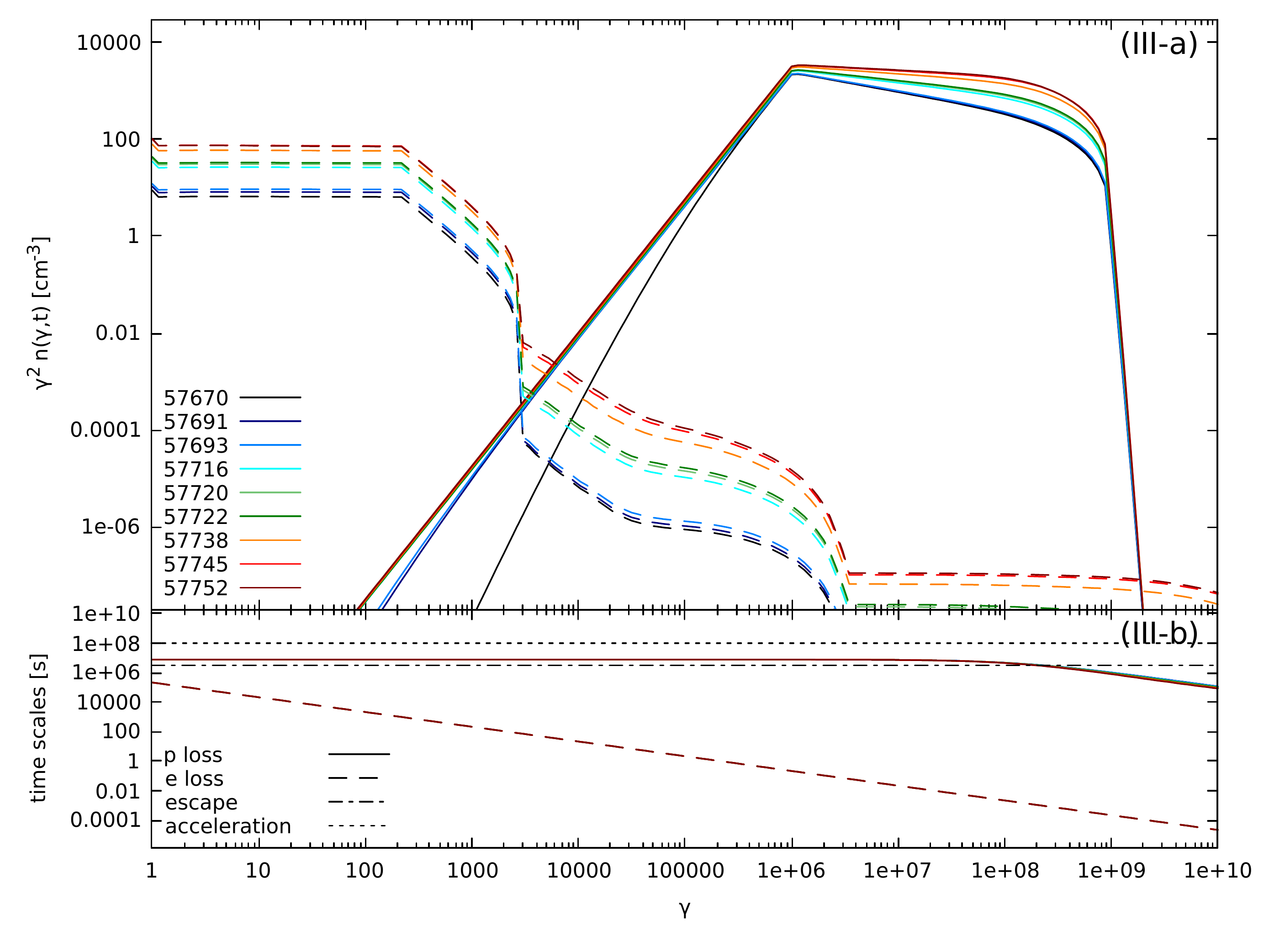}}
\end{minipage}
\hspace{\fill}
\begin{minipage}{0.49\linewidth}
\centering \resizebox{\hsize}{!}
{\includegraphics{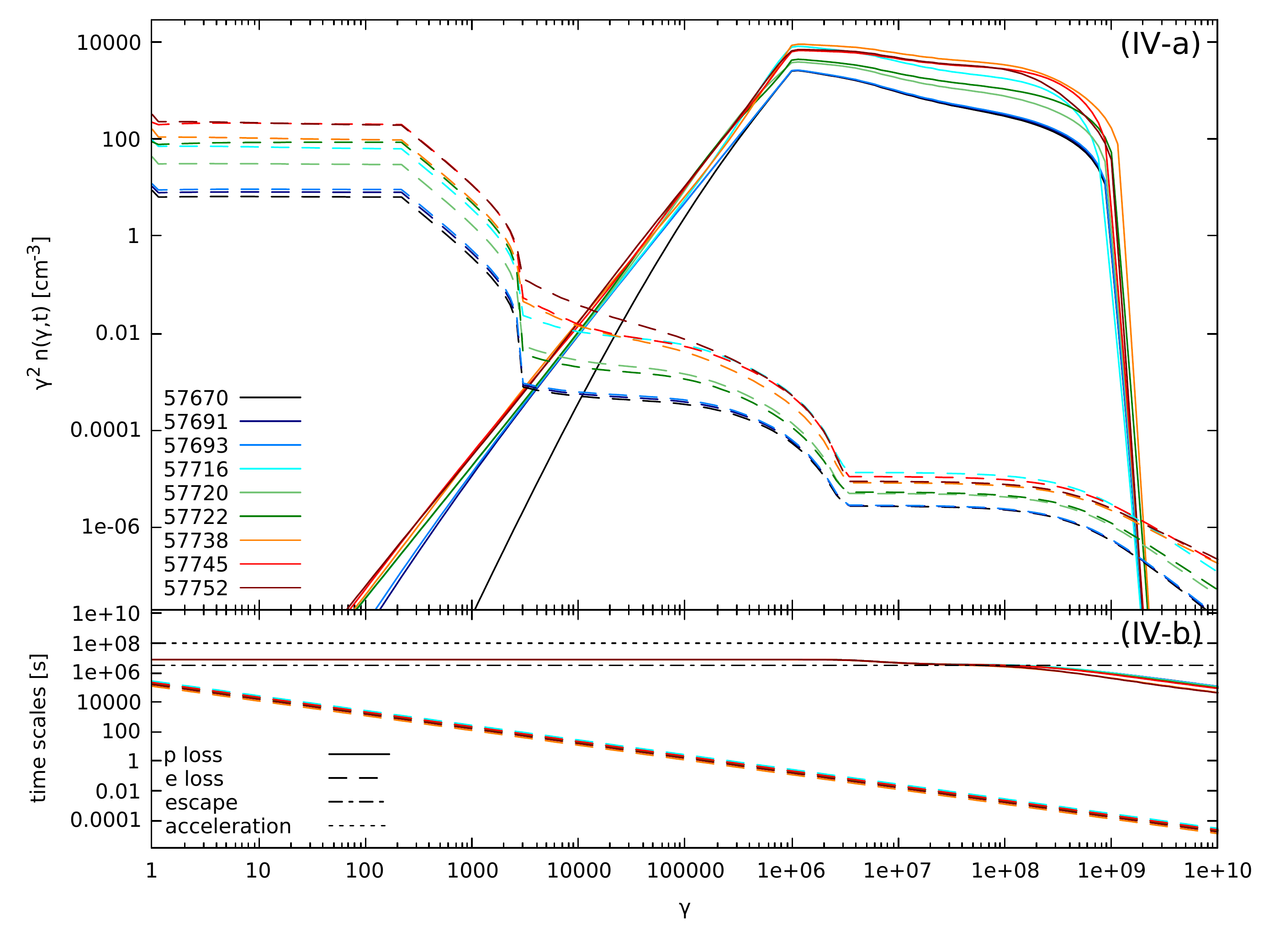}}
\end{minipage}
\caption{{\bf Panel I:} {\it (a)} Proton (solid) and electron (dashed) distribution functions $\gamma^2 n_i(\gamma,t)$ as a function of the particle Lorentz factor $\gamma$ for the same time steps as in Fig. \ref{fig:mod-spec} for an emission region at $z=6.5\E{17}\,$cm. 
{\it (b)} Proton (solid) and electron (dashed) cooling times $\gamma/|\dot{\gamma}_i|$, as well as the escape time (dotted) and the acceleration time (dash-dotted) as a function of the particle Lorentz factor $\gamma$.
{\bf Panel II:} Same as in panel I, but for an emission region at $z=3.09\E{18}\,$cm.
{\bf Panel III:} Same as in panel I, but for an emission region at $z=3.09\E{19}\,$cm.
{\bf Panel IV:} Same as in panel I, but for an emission region at $z=3.09\E{18}\,$cm and the subflares \change{(cf. appendix \ref{sec:modIV})}.
}
\label{fig:mod-part-I}
\end{figure*} 
%
% MB: In the lower panels the different line styles are indistinguishable
% MZ: done.
%
\begin{table*}[th]
\caption{Parameters used in individual examples. Given are the parameter description, symbol and value. The parameters below the horizontal line describe the variability.}
\begin{tabular}{lcl|c|c|c}
Definition							& \multicolumn{2}{c|}{Symbol} 		& Example I				& Example II			& Example III  \\
\hline
Emission region distance			& $z$					& [cm]		& $6.50\times 10^{17}$	& $3.09\times 10^{18}$	& $3.09\times 10^{19}$ \\
Proton injection luminosity			& $L_{\rm p,inj}$		& [erg/s]	& $2.2\times 10^{44}$	& $1.3\times 10^{44}$	& $1.1\times 10^{44}$ \\
Electron injection luminosity		& $L_{\rm e,inj}$		& [erg/s]	& $3.8\times 10^{41}$	& $3.2\times 10^{41}$	& $3.2\times 10^{41}$ \\
\hline
Proton injection variability		& $\mathscr{L}_{\rm p}$	& [erg/s]	& $8.3\times 10^{43}$	& $5.0\times 10^{43}$	& $4.8\times 10^{43}$ \\
Proton spectral index variability	& $\mathscr{s}_{\rm p}$	& 			& $-0.3$ 				& $-0.3$			  	 & $-0.3$  \\
Electron injection variability		& $\mathscr{L}_{\rm e}$	& [erg/s]	& $9.5\times 10^{41}$	& $8.0\times 10^{41}$	& $8.0\times 10^{41}$ 
\end{tabular}
\label{tab:modelparam}
\end{table*}
%
% JPL: explain the $\mathscr{s}_{\rm p}$ parameter ?
% MZ: Good point, thanks.
%
The leptonic model in paper I required the BLR photons as the target for the inverse Compton process to produce the observed amount of \g-rays. If the emission region would be located beyond the BLR, the leptonic one-zone model cannot reproduce the observations. This is different for hadronic one-zone models, where the bulk of \g-rays is produced through proton synchrotron. It is therefore possible to place the emission region at different distances $z$ from the black hole and we explore three possibilities, namely at the outer edge of the BLR as in paper I, within the DT and outside the DT. The parameters for these examples are given in Tabs.~\ref{tab:genericparam} and \ref{tab:modelparam}.
The different emission region distances have consequences for the model of the incoming cloud. We assume that the cloud orbits the supermassive black hole. Hence, at larger distances from the black hole, the orbital velocity is smaller. Since the radius of the cloud is determined by both the velocity and the (constant) duration of the long-term event, the resulting radius of the cloud decreases for larger distances from the black hole. The resulting changes in the particle density of the cloud give different estimates of the cloud's temperature. The equations to calculate the cloud's velocity, radius, density and temperature are given in Eqs. (\ref{eq:vc}), (\ref{eq:Rc}), (\ref{eq:clouddens}), and (\ref{eq:cloudtemp}), respectively.

The cloud ablation model of paper I provides the particle injection luminosity, cf. Eq.~(\ref{eq:injection}), as
\begin{align}
L_{\rm i,inj}(t) = L_{\rm i,inj} + \mathscr{L}_i \ln{\left( \frac{t_0^2+t_c^2}{t_0^2+(t_c-t)^2} \right)}, \label{eq:cloudinj}
\end{align}
where the index $i$ is either protons (p) or electrons (e), $L_{\rm inj}$ is the steady-state particle injection power, $\mathscr{L}$ is the particle injection power of the variability, $t_c^{\rm obs}$ is the observed time from the beginning to the peak of the long-term event giving $t_c = \delta t_c^{\rm obs}/(1+\zred)$, and $t_0 = t_c/8.3$. The observed time scale of the flare $t_c^{\rm obs}$ is related to the radius of the ablated gas cloud, while $t_0$ is related to the cloud's scale height. The scale height is a free parameter in our model, and the given ratio of $8.3$ provides the best fit to the data. The influence of the scale height on the form of the lightcurve will be discussed elsewhere.

Along with the variation in the particle power according to Eq.~(\ref{eq:cloudinj}), we vary the spectral index of the proton distribution as discussed in section~\ref{sec:cons}. We assume a linear change of the spectral index with time,
\begin{align}
s_{\rm p}(t) = s_{\rm p} + \mathscr{s}_{\rm p} \frac{t_c-|t_c-t|}{t_c} \label{eq:protonvar},
\end{align}
with $\mathscr{s}_{\rm p}$ being the variation of the proton spectral index. While the linear variation does not have a physical motivation, it gives a satisfactory representation of the event and reproduces well the constraints of Tab.~\ref{tab:paramconst}.

Below, we present the results for the three different distances of the emission region of the black hole.%, followed by a brief discussion about their implications. 
\change{We also consider a few of the medium-term subflares for which we have a detailed data set, as shown in Figs.~\ref{fig:mod-lc}, \ref{fig:mod-spec}, and \ref{fig:mod-part-I}. Details of the subflares and their parameters are presented in appendix \ref{sec:modIV}.}

%###############################################################################################################################
\subsubsection{Example I: edge of the broad-line region} \label{sec:modI}
In this setup the emission region is placed at the outer edge of the BLR at $z=6.5\times 10^{17}\,$cm -- at the same position as in paper I. Only the ablation of the gas cloud is considered, which provides a description of the overall trend but does not account for subflares. The solid lines in Figs.~\ref{fig:mod-lc} and \ref{fig:mod-spec} show the resulting light curves and SEDs, respectively. 

As for the leptonic case of paper I, the hadronic model is able to reproduce the long-term evolution of the event. The \g-ray, R-band and V-band light curves are fit very well, where the model curves follow the lower boundary of the data points. The X-ray light curve is not fit very well, even though the general behavior of only a mild increase (ignoring the subflares) is reproduced.

The position near the outer edge of the BLR limits the absorption of the \g-rays, which implies that the emission of a $\sim 200\,$GeV photon (in the frame of \cta), as detected with \fermi, is possible. The model photon and neutrino spectra are shown as thick and thin solid lines in Fig.~\ref{fig:mod-spec}, respectively, for the 9 dates of contemporaneous spectra. Given that many of these spectra were obtained during subflare episodes, the model is not expected to reproduce the data perfectly. %However, the general agreement is okay.

Fig.~\ref{fig:mod-part-I}(I) shows the particle distributions and the cooling time scales for the spectra shown in Fig.~\ref{fig:mod-spec}. The dashed lines in panel (a) show the electron distributions, which exhibit the expected broken power-law at low energies, and the secondary population from muon decay at high energies \change{(the distinction is made at the maximum injection Lorentz factor $\gamma_{\rm e,max}=3\E{3}$)}. As can be seen in panel (b) -- which shows the important time scales for the particle distributions -- the electron cooling is solely governed by synchrotron cooling, and no contribution from the external field can be seen -- which would reveal themselves by a parallel shift to shorter times at low energies, c.f. paper I. The proton distribution (solid lines) also resembles a broken power-law with an exponential decay at the highest energies. The dip seen after the break stems from a change in the cooling behavior. At the highest energies the proton cooling is dominated by synchrotron cooling in the fast cooling regime (i.e., cooling time is shorter than the escape time), and then switches to cooling through pion production -- still in the fast cooling regime. At energies below the pion production threshold (corresponding in this case to a proton Lorentz factor of $\gamma\sim 10^6$) the cooling switches to adiabatic cooling in the slow cooling regime. The latter change explains the dip in the particle distribution at the same energy. This dip in the particle distribution is responsible for the dip seen in the SED above $10\,$keV.

%###############################################################################################################################
\subsubsection{Example II: within the dusty torus} \label{sec:modII}
In this example, the emission region is placed at $z=1\,$pc from the black hole. This is far away from the BLR, but within the DT, reducing the energy density of the external photon field. The dashed lines in Figs.~\ref{fig:mod-lc} and \ref{fig:mod-spec} show the model lightcurve and spectra, respectively. The overall agreement is slightly better than in Example I.

The reason for the improved fit can be seen in Fig.~\ref{fig:mod-part-I}(II), which displays the particle distributions and time scales for this setup. The dip in the proton distribution due to pion production is also present, but it is less pronounced and at higher energies ($\gamma\gtrsim 10^7$) than in Example I. The pion production threshold depends on the proton Lorentz factor and the incident photon energy. Since the average energy of the DT photons is lower than the average energy of BLR photons, the pion production threshold moves to higher proton Lorentz factors compared to Example I. This can be clearly identified in panel (b) through the proton cooling time scale, where the change from slow to fast cooling owing to cooling through pion production has shifted to higher energies. This leads to a smoother particle distribution, and a smoother proton synchrotron component between the X-ray and \g-ray SED points. A consequence of the smoother particle distribution is a reduction in the proton number for the same radiative power output. Hence, the reduction in the proton injection power (cf. Tab.~\ref{tab:modelparam}). 
The smoother particle spectrum explains the improved fit of the X-ray lightcurve and spectrum.

The reduced pion production leads to fewer secondary electrons and positrons, resulting in a reduction of high-energy electrons above the injection spectrum. The electron cooling is unchanged owing to the strong dominance of the synchrotron process.

%###############################################################################################################################
\subsubsection{Example III: outside the dusty torus} \label{sec:modIII}
Located at $z=10\,$pc from the black hole, the emission region is beyond the photon fields of the BLR and DT. Thus, pion production in this example depends only on the internal synchrotron fields, and is much reduced. Hence, proton cooling is solely determined by adiabatic losses at most energies and synchrotron cooling at the highest energies ($\gamma>10^8$). This also means that the proton distribution is smooth without any cooling ``dips'', and hence the emerging synchrotron flux is a single smooth power-law between the observed X-ray and \g-ray energies. Since this requires even less protons than for Example II, the X-ray flux is further reduced, explaining the slight underrepresentation of the X-ray lightcurve in Fig.~\ref{fig:mod-lc}. The low pion production rate further results in a low number of secondary electrons and positrons, as can be seen in Fig.~\ref{fig:mod-part-I}(III).

%###############################################################################################################################
%\subsubsection{Comparison of Examples I-III} \label{sec:compI-III}
%

%
%################################################################################################################################
\section{Discussion} \label{sec:dis}
%
% - use energy density comparison between cloud and jet to argue while most of the material will not enter the jet !!!DONE!!!
% - Where is the more likely position of the emission region? !!!DONE!!!
% - What if it is an entire star forming region (aka Fermi long-term light curve!)? !!!DONE!!!
% - Neutrinos !!!DONE!!!
% - fast flares !!!DONE!!!
% - SYMMETRY of the long light curve !!!DONE!!!
%
In this section, we discuss the implications and results of the afore described modelings. We start with \change{the jet power, followed by the} neutrino output predicted by our model, since it might serve as a potential discriminator between hadronic and leptonic models. Thirdly, we discuss how the fast flares observed by us and others can be incorporated into the framework of our model. Lastly, we discuss \change{the inferred parameters of} the cloud. %For that we also consider the ``bigger picture'' in terms of the data accumulated \change{since mid-2008} with \fermi, {\it Swift} and ATOM. \change{In this section we also refer to the results of Example IV (where we include a few of the subflares). Details are given in appendix \ref{sec:modIV}.}

\change{
\subsection{Jet power} \label{sec:jetpower}
}
\begin{table*}[th]
\caption{Summary of the modeling results. The Eddington power of CTA\,102's black hole is $P\p_{\rm Edd}=1.1\E{47}\,$erg/s. Parameters below the horizontal line give the inferred cloud parameters.}
\begin{tabular}{lcl|c|c|c}
Definition		& \multicolumn{2}{c|}{Symbol} 						& Example I		& Example II 	& Example III  \\
\hline
Minimum proton power		& $P\obs_{\rm p,min}$	& [erg/s]		& $4.1\E{47}$	& $4.6\E{47}$	& $4.2\E{47}$ \\
Maximum proton power		& $P\obs_{\rm p,max}$	& [erg/s]		& $9.5\E{47}$	& $1.1\E{48}$	& $1.1\E{48}$ \\
Minimum electron power		& $P\obs_{\rm e,min}$	& [erg/s]		& $1.4\E{42}$	& $1.5\E{42}$	& $1.5\E{42}$ \\
Maximum electron power		& $P\obs_{\rm e,max}$	& [erg/s]		& $1.6\E{43}$	& $1.7\E{43}$	& $1.7\E{43}$ \\
Minimum radiative power		& $P\obs_{\rm r,min}$	& [erg/s]		& $7.8\E{45}$	& $7.0\E{45}$	& $7.0\E{45}$ \\
Maximum radiative power		& $P\obs_{\rm r,max}$	& [erg/s]		& $4.4\E{46}$	& $4.1\E{46}$	& $4.3\E{46}$ \\
%Minimum magnetic power		& $P\obs_{\rm B,min}$	& [erg/s]		& $6.6\E{48}$	& $6.6\E{48}$	& $6.6\E{48}$ \\
Magnetic power				& $P\obs_{\rm B,max}$	& [erg/s]		& $6.6\E{48}$	& $6.6\E{48}$	& $6.6\E{48}$ \\
Minimum proton density		& $n_{\rm p,min}$		& [cm$^{-3}$]	& $5.9\E{-3}$	& $3.5\E{-3}$	& $3.0\E{-3}$ \\
Maximum proton density		& $n_{\rm p,max}$		& [cm$^{-3}$]	& $9.1\E{-3}$	& $5.4\E{-3}$	& $5.0\E{-3}$ \\
Minimum electron density	& $n_{\rm e,min}$		& [cm$^{-3}$]	& $6.2\E{0}$	& $6.7\E{0}$	& $6.7\E{0}$ \\
Maximum electron density	& $n_{\rm e,max}$		& [cm$^{-3}$]	& $7.3\E{1}$	& $7.9\E{1}$	& $8.0\E{1}$ \\
\hline
Cloud speed			& $v_{\rm c}$	& [cm/s]		& $4.2\E{8}$	& $1.9\E{8}$	& $6.1\E{7}$ \\ 
Cloud radius		& $R_{\rm c}$	& [cm]			& $1.1\E{15}$	& $4.9\E{14}$	& $1.5\E{14}$ \\ 
Cloud density		& $n_{\rm c}$	& [cm$^{-3}$]	& $9.5\E{5}$	& $1.1\E{7}$	& $3.4\E{8}$ \\ 
Cloud mass			& $M_{\rm c}$	& [g]			& $8.9\E{27}$	& $9.1\E{27}$	& $8.0\E{27}$ \\ 
Cloud temperature	& $T_{\rm c}$	& [K]			& $1.1\E{-3}$	& $2.7\E{-3}$	& $8.7\E{-3}$ 
\end{tabular}
\label{tab:modelres}
\end{table*}
The powers in the individual jet constituents (particles, radiation and magnetic field) are listed in Tab.~\ref{tab:modelres}. Both minimum and maximum values are given, where the former corresponds to the period before the event started ($\sim$MJD~57690), and the latter corresponds to the time of the maximum of the light curve ($\sim$MJD~57750). The magnetic power does not change. It is the dominating constituent in terms of power even during the time of the maximum flux \change{dominating the protons by at least a factor 6}. 

In Example I, the proton power changes by about a factor $2.3$ between the onset and the maximum of the outburst. Both the magnetic and the proton power exceed the Eddington power of the black hole ($P\p_{\rm Edd}=1.1\E{47}\,$erg/s) even before the onset of the flare. This is a characteristic \citep[e.g.,][]{zb15}, yet unsolved problem of hadronic models. \change{While several disk models allow for super-Eddington accretion, it is at least questionable if these states can last on long time scales. Naturally, some of the jet power could also be extracted from the rotation of the black hole \citep{bz77}. Whether these effects combined can support a super-Eddington jet on long time scales, is an interesting questions that must remain unanswered for now.} The radiative power, and even more so the electron power are subdominant.

While the number of protons is slightly reduced in Example II compared to Example I, the overall power in the proton population has increased. Since the cooling of protons is less efficient at highest energies than in Example I, a larger fraction of protons exhibits higher energies explaining the larger power despite the slight decrease in total numbers. Apart from that we find the same results as in Example I: The jet power exceeds the Eddington power of the black hole, and is dominated by the magnetic field and the proton power, while the radiative and electron powers are minor.

For the powers of the jet constituents in Example III, the same statements hold as in the other cases. Proton and magnetic field powers dominate by far the radiative and electron powers, giving a total power exceeding the Eddington power of the black hole.

\subsection{Neutrino emission} \label{sec:neu}
A by-product of hadronic interactions through pion and muon decay is neutrino production. Hence, one of the most important tests to make is the detection of neutrinos from the direction of an active galaxy. This would directly quantify hadronic processes. A hint of such linked neutrino and electromagnetic emission from an AGN has recently been reported for TXS\,0506$+$056 \citep{aarst18}.
% MB: I would drop "esp." since neutrinos are produced ONLY through pion and muon decay
% MZ: Done.

Our code allows us to calculate the neutrino spectrum and production rate folded with the IceCube effective area  \citep{icc13}. A subsequent integration of the production rate over all time steps gives the total number of detectable neutrinos for IceCube in that time interval. The long-term flare is confined between MJD~57685 and MJD~57815. The number of detectable neutrinos in that time interval is for Example I $4.9\E{-3}$ neutrinos, for Example II $1.1\E{-3}$ neutrinos, for Example III $2.5\E{-5}$ neutrinos, and for Example IV $1.6\E{-3}$ neutrinos. All these numbers are significantly smaller than unity, and so the odds are low that IceCube has detected a neutrino from \cta. 

Nonetheless, it is interesting to see the difference in these numbers caused by the different amount of external photons penetrating the emission region. Compared to (almost) no external photons in Example III, the dusty torus increases the amount of neutrinos by about a factor 40 (Example II), while the BLR enhances the neutrino number by more than a factor 200 compared to no external photons (Example I). Hence, while the photon light curves in these 3 baseline models are basically unchanged (at least in the HE \g-ray and optical domain, cf. Fig.~\ref{fig:mod-lc}), the neutrino emission is significantly influenced. This could become a powerful asset, once a more sensitive neutrino detector is built, to not just judge on the underlying emission model, but also to put constraints on the emission region location.
% MB: Add an "external" to photons in line 3
% MZ: Done.

The addition of the subflares only increases the number of neutrinos by about 50\% compared to \change{Example II, which is the baseline model in Example IV}. Hence, while the subflares raise the number of neutrinos, they would not inhibit the location constraint owing to the large difference in neutrino numbers expected from the different locations.

The neutrino SEDs, shown in Fig.~\ref{fig:mod-spec} as thin lines, also exhibit remarkable differences. The external photon fields lead to both a significant flux increase and a significant shift of the peak energy to lower energies. The peak energy drops by more than 2 orders of magnitude between Examples III and I (from about $100\,$PeV to $1\,$PeV). The subflares in Example IV cause secondary peaks at about 100\,PeV (see the lower row of panels in Fig.~\ref{fig:mod-spec}), which are solely due to the change in the source internal photon fields, since they are at the same energy as the peak of Example III. However, the detection of these differences in the neutrino spectra will require not just the detection of some neutrinos, but of a large number. Unfortunately, this might not be possible with next-generation detectors.

The difference in neutrino fluxes depending on the different Examples is not only due to a different number in soft photons, but also their different energy densities. Since the pion production threshold is fixed to a center-of-mass energy of $\sqrt{s}\sim 1.07\,$GeV, the lower the soft photon energy, the higher the required proton energy. Hence, many more pions will be produced within the BLR than in other regions owing to both a larger number of soft photons and to the larger number of protons (due to the lower energy required by protons) being available for the interaction. This also explains the different spectral forms, since the neutrinos take a fixed energy of the initial proton.
% MB: It might be worthwhile to mention explicitly the reason for the different neutrino rates: "Pion-Production threshold -> lower-energy photons require higher-energy protons to make pions -> neutrinos take up a fixed percentage of the proton energy ..."
% Done.

%##################################################
\subsection{The fast flares} \label{sec:fastflares}
Our modeling aimed for an explanation of the long-term trend, and variability that lasted more than a day, like the six subflares in Example IV. We purposely neglected the short-term variability on the order of hours and minutes that is present in our data and that was also reported by others. Nonetheless, the presence of the short-term variability has an important consequence.

Within the one-zone model many authors use the implicit assumption that the emission region fills the entire cross-section of the jet. This is used to estimate the distance of the emission region from the black hole assuming a constant opening angle of the jet:
\begin{align}
 z = \frac{R}{\tan{(a/\Gamma_b)}} \approx \frac{R\Gamma_b}{a} \label{eq:distest},
\end{align}
where $\Gamma_b$ is the bulk Lorentz factor, which we assume to be equal to the Doppler factor $\delta$, and $a$ being a scaling parameter for the opening angle (typically $a\sim 1$).

In the modeling we have used $R=2\E{16}\,$cm, which corresponds to a variability time scale of about half-a-day, Eq.~(\ref{eq:tvar}). Using the distance $z=6.5\E{17}\,$cm from Example I, which is also constrained well as the minimum distance, c.f. the discussion in section~\ref{sec:blr}, the scaling parameter becomes $a\sim 1.1$. In this case, the emission region can fill the cross-section of the jet. 

Using instead the minimum variability present in our data, which is $0.03\,$d from R-band observations with ATOM, the emission region size cannot be larger than $R\sim 1.3\E{15}\,$cm. For the emission region to be located the same distance from the black hole, $a\sim 0.07$, which would be an extremely small opening angle of the jet. Using the minute-scale variability of the \g-ray light curve \citep{sea18} would reduce this estimate by another factor 5 or so. It is, thus, much more likely that the fast flares originate from substructures within a larger emission region. This is in line with findings on other FSRQs, where fast variability has been observed and the emission region can be confidently placed on the outer edge or beyond the broad-line region \citep{Rea18,Zea18}.

Incorporating such fast flares within a hadronic scenario is challenging, since the cooling time scale of protons is much longer than that of electrons (see Fig.~\ref{fig:mod-part-I}). A small emission region implies a fast escape of particles without significant cooling, and therefore a very low radiative efficiency. As discussed by \cite{pea17}, small emission regions with kG magnetic fields would be able to account for fast and bright flares within a hadronic model. This could be in line with substructure within the main emission region \citep{g13} or a turbulent multi-zone model \citep{m14}.

Hence, the addition of small zones within the emission region in the jet could account for the fast flares within the general framework outlined in this paper. These could originate from turbulence and/or magnetic reconnection. If the gas cloud is magnetized, reconnection events between the magnetic fields of the cloud and the jet could lead to these fast accelerations.

\change{
%##################################################
\subsection{Cloud parameters} \label{sec:cloud}
}
The particle densities in Tab.~\ref{tab:modelres} allow us to calculate the number of particles that need to be provided by the cloud, which we can then use to obtain the parameters of the cloud (see appendix \ref{app:theory}). The resulting cloud parameters are also listed in Tab.~\ref{tab:modelres}. 

In case of Example I, the emission region is placed at the same distance from the black hole as in paper I. Hence, the velocity and radius of the cloud are also the same. However, since the hadronic model requires a larger magnetic field in the emission region than the leptonic one, in turn it does not require as many particles for the same radiative output. Hence, the inferred density of the cloud is low, and so is the temperature.

The emission region in Example II is further away from the black hole than in Example I. Hence, the cloud moves slower. Given the constant duration of the long-term event, we infer a smaller cloud radius, and a larger density and temperature than in Example I. 
In Example III, the speed is slowed down even more, reducing the radius and increasing the density and temperature compared to the other cases.

\change{The speeds are determined from the Keplerian velocity at a given distance from the black hole. The speed in Example III is on the order of $600\,$km/s. Further out, the influence of the black hole is diminished and the speed is determined by the combined gravitational field of the whole galaxy. In elliptical galaxies the typical dispersion speed is $\sim 300\,$km/s in the inner kpc. Individual speeds would therefore be on the order of $\sim 100\,$km/s, about a factor 6 less than what we use in Example III. Using this speed, we would obtain a cloud radius of $R_c=2.6\E{13}\,$cm, a density of $n_c=3.6\E{10}\,$cm$^{-3}$, and a temperature $T_c=50\,$mK.
}

The previous discussion was done under the implicit assumption that the entire material intercepted by the jet is also devoured by it \change{and accelerated into the non-thermal particle spectrum}. As in paper I, the inferred temperature of the gas cloud is much below the temperature of the CMB. \change{While there are apparently clouds in space that can be colder than the CMB \citep[e.g. the Boomerang Nebula,][]{sea13}, temperatures on the order of milli-Kelvin are unlikely.}
The temperature estimate would change substantially, if only \change{a fraction of the devoured particles are accelerated (as, e.g., in supernova shocks, where less than $10\%$ of the particles are accelerated), and if only parts of the cloud actually} enter the jet and most material is ejected during the crossing. We calculate the energy densities present in both the jet and the cloud before the interaction. As a proxy for the jet energy density and ram pressure, we use the magnetic energy density of the emission region $u_B = 143\,$erg/cm$^3$, \change{as this is the dominating entity (see Tab.~\ref{tab:modelres})}. One should note that this is possibly only a lower limit on the magnetic pressure. The mechanism to confine relativistic jets might involve strong magnetic fields, which would increase the magnetic energy density at the jet boundary. The thermal pressure of the cloud can be used as a proxy for the energy density in the incoming cloud, giving 
\begin{align}
 u_{th} = 1.4\E{-5} \est{n_c}{1.0\E{10}\,\mbox{cm}^{-3}}{} \est{T_c}{10\,\mbox{K}}{} \, \mbox{erg/cm}^3 \label{eq:uthermcloud}.
\end{align}
Using parameters for the larger clouds given in Tab.~\ref{tab:modelres}, we find that the thermal energy density for the cloud of Example II is $u_{th} = 3.7\E{-12}\,$erg/cm$^3$, which is almost 14 orders of magnitude below the magnetic energy density of the jet. 
Dense clouds in interstellar space typically exhibit temperatures of $\sim 20\,$K. Using the spatial dimensions of the Example II cloud and Eq.~(\ref{eq:cloudtemp}), we can calculate the density of the cloud for a temperature of $20\,$K, which is $n_c\sim 4\E{10}\,$cm$^{-3}$. The thermal energy density in this case is $u_{th}\sim 1\E{-4}\,$erg/cm$^3$. This is still 6 orders of magnitude below the magnetic energy density of the jet. This huge difference in the energy densities implies that the jet will look like a giant wall to the cloud, and most of the cloud's particles could be reflected during the crossing and would not enter the jet. Statistically, the fraction of particles entering the jet would follow the particle distribution within the cloud, and the injection of the type of Eq.~(\ref{eq:injection}) would still be applicable. This could explain why the estimates of the temperature of the cloud give unrealistic numbers.

\change{
%
%################################################################################################################################
\section{Nature of the cloud} \label{sec:origin}
}
\begin{figure*}[th]
\centering
\includegraphics[width=1.00\textwidth]{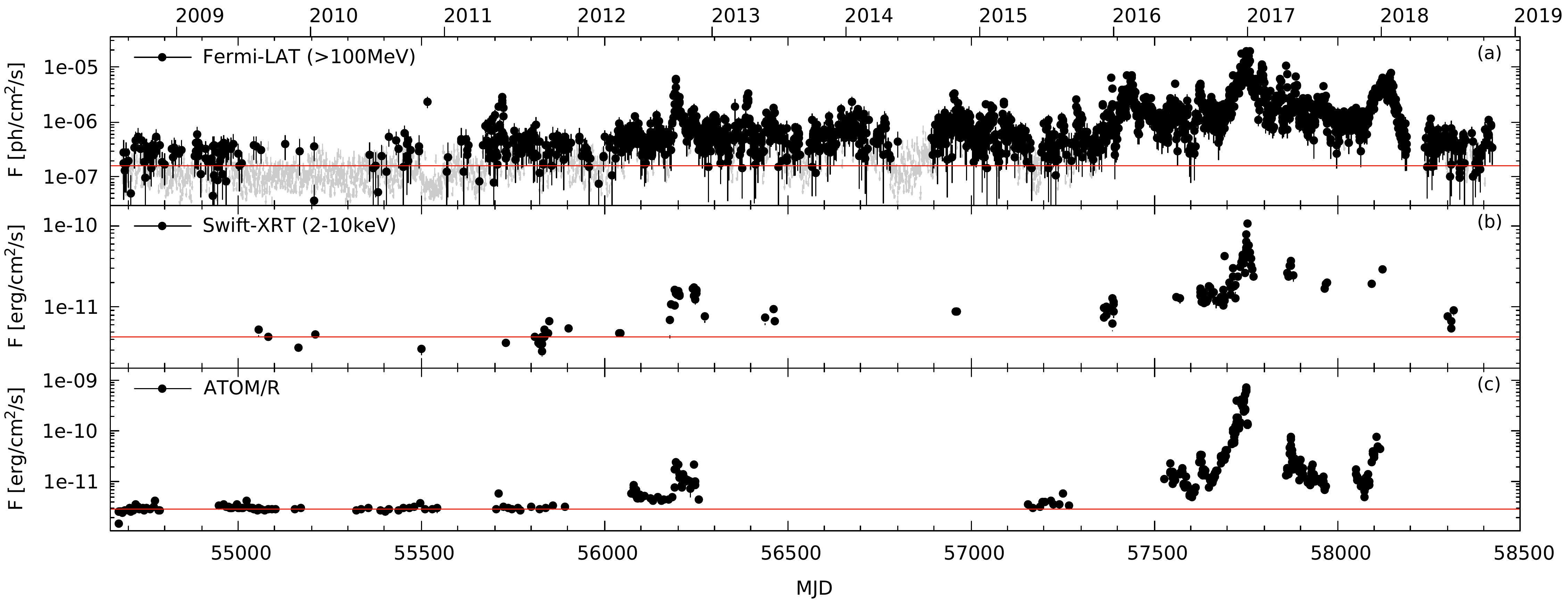}
\caption{Long-term light curves of CTA\,102. In all panels, the red line marks the average before 2012. Note the logarithmic scale on the y-axis.  
{\bf (a)} Daily HE \g-ray light curve observed with \fermi. The gray arrows mark upper limits.
{\bf (b)} {\it Swift}-XRT light curve for individual pointings.
{\bf (c)} ATOM R-band light curve for individual pointings.
%\\ {\bf UPDATE WITH MOST RECENT LIGHTCURVE POSSIBLE BEFORE SUBMISSION!!!!!}
}
\label{fig:ltlc}
\end{figure*} 
So far, we have not considered the nature of the cloud causing the flare. Cloud-like structures are proposed in different regions within AGN \change{and the host}. The most obvious choice close to the black hole are BLR clouds. They are typically given with a radius of $\sim 10^{13}\,$cm and densities of $10^{9..11}\,$cm$^{-3}$ \citep{dea99,p06}. This fits well with the derived parameters of the clouds responsible for the subflares, as derived in appendix \ref{sec:modIV} and listed in Tab.~\ref{tab:modelresIV}.\footnote{\change{Note that we continue to use the parameters given in Tabs.~\ref{tab:modelres} and \ref{tab:modelresIV}. Following the discussion in section \ref{sec:cloud} these values are probably lower limits. However, how much denser the clouds might be cannot be quantified.}} 
However, the main flare cannot be reproduced by a BLR cloud with such parameters. 

Cloud-like structures are also seen within star-forming regions. Embedded in giant molecular clouds, the individual star forming cloud cores can reach sizes of $0.05-1\,$pc and densities of $10^{7..9}\,$cm$^{-3}$ \citep[e.g.,][ch. 12]{co17}. While too large compared to the derived cloud radii of Examples I-III, the densities fit quite well. On the other hand, the density is an average over the core, which hosts several forming stars. Hence, the actual sites of star formation are smaller, and might therefore work as a seed for this flare, as well. The subflares might then be a product of density fluctuations within the cloud or forming stars themselves. These should have been distributed throughout the cloud core resulting in numerous subflares, which is indeed the case. 

A different possibility for the nature of the cloud might be the atmosphere of a red giant star (RG). RGs are post main-sequence stars with highly inflated atmospheres with a density structure very similar to Eq.~(\ref{eq:ncloud}). \cite{bab10} and others considered such a model for day-long flares in M\,87 assuming that the RG first fully enters the jet before the ablation process begins. As we have shown in paper I, the jet ram pressure can easily ablate material even from the surface of a star. Hence, the process probably begins immediately. The total masses of the clouds given in Tabs.~\ref{tab:modelres} and \ref{tab:modelresIV} are on the order of $10^{28}\,$g, which is significantly less than the mass of the sun. Atmospheres of RGs can reach radii of $10^{13}\,$cm. This is 1 or 2 orders of magnitude too small for the long-term event, but fits well the estimated radii of the subflare regions. On the other hand, these radii depend strongly on the velocity of the cloud (or RG). If the object would be moving slower \change{than what we considered in the Examples, the radius would shrink. In fact, the radius would fit nicely for a cloud moving at $\sim 100\,$km/s, as mentioned in section \ref{sec:cloud}.} Hence, the interaction of an RG with the jet at large distances from the black hole is a plausible scenario.

%This is different for BLR clouds and RGs. These objects exhibit temperatures on the order of $10^4\,$K. In this case the energy density of these objects reaches $u_{th}\sim 1\E{2}\,$erg/cm$^3$, which is on the same order of magnitude as the jet's magnetic energy density.

%We can repeat the exercise for all the other clouds in the above discussion, for which the same arguments can be made. If in all models the cloud densities are actually at least 3 or more orders of magnitudes larger than the estimates, the models I - III imply densities of $10^{9..11}\,$cm$^{-3}$ in the large clouds, while the subflares reach $10^{13..15}\,$cm$^{-3}$. All these values depend, of course, on the speed of the respective clouds and therefore on the distance from the black hole. Our models fit roughly equally well -- so, we cannot distinguish on the modeling and this data set alone. 

%With these densities, the larger cloud exhibits densities typical of the BLR. In that scenario, the subflares might be a consequence of instabilities triggered through the collision or the injection process. A star forming region is on average not thick enough to account for such clouds -- even though, denser cores within the star forming region are a possibility. 

%While this unfortunately cannot become more than speculative statements, there is a certain appeal to the idea that the flare has been caused by a star-forming region. 
The 2016-2017 giant outburst of \cta\ is part of a period lasting several years of tremendous activity. In Fig.~\ref{fig:ltlc}(a) we show the HE \g-ray light curve of \cta\ observed since the launch of \fermi. While the entire light curve shows a large number of flares, the average or low state remained roughly constant at a level of $\sim 1.6\E{-7}\,$ph/cm$^2$/s (cf. the red line in Fig.~\ref{fig:ltlc}(a), which marks the flux quoted in the 3FGL catalog) until a strong outburst in early 2016 (MJD~$\sim 57400$). Since then the flux has exceeded this level continuously by about a factor 5. The giant outburst of 2016-2017 is prominently visible. After that the source level remained enhanced until another strong flare in early 2018 (MJD~$\sim 58200$), after which the source has become much quieter, seemingly returning to pre-2016 flux levels. While the cadence of observations by {\it Swift} and ATOM is much lower than that of {\it Fermi}, similar statements can be made. As shown in Fig.~\ref{fig:ltlc}(b) and (c), respectively, the X-ray and optical fluxes were also enhanced for an extended period of time between 2016 and 2018 compared to the average states integrated for date before 2012 ($\sim 4.3\E{-12}\,$erg/cm$^2$/s for the X-ray lightcurve and $\sim 2.9\E{-12}\,$erg/cm$^2$/s for the optical lightcurve). This is an important point, since in a hadronic model the optical lightcurve samples the electron variation and the X- and \g-ray lightcurves sample the proton variation. The apparent simultaneous change in the lightcurves from 2016 to 2018 implies that the number of electrons and protons varied similarly, and could have been fed from a common plasma reservoir.

It is tempting to assume that the giant flare discussed in detail in this paper could be part of a much longer event that lasted about 800\,d. The fact that the fluxes never returned to the old state during a period of more than two years along with the fascinating symmetry of the lightcurves during this 800\,d period (similar to the symmetry of the giant flare, which is right in the middle of the 2-year period) point towards a common origin. Whether this is indeed the case or just a lucky coincidence is difficult to say. A modeling of the 800\,d lightcurve might give a hint depending on the resulting parameters of the ablated object. This is, however, beyond the scope of this paper.
\section{Summary} \label{sec:sum}
In this paper, we have revisited the cloud ablation model of paper I to explain the giant, 4-months long outburst of \cta\ in late 2016 to early 2017. In paper I we used a leptonic radiation model to reproduce the outburst, which let to the requirement that the emission region needed to be placed at the outer edge of the BLR. Here we use a hadronic model. While it results in super-Eddington jet powers at all times, which is a common \citep[e.g.,][]{zb15} yet unsolved problem for hadronic models, it allows us to place the emission region at different locations along the jet, since the bulk of \g-ray flux is reproduced with proton synchrotron emission. The reason to explore the different locations is that the inferred parameters of the incoming cloud in paper I could not be matched well with parameters of specific objects, such as BLR clouds. Since the duration of the event is fixed, the radius and density of the cloud depend strongly on the cloud's velocity, which in turn depends on the distance from the black hole.

We have explored three different examples: The interaction region being at the outer edge of the BLR (as in paper I) at $z=0.2\,$pc, within the DT at $z=1\,$pc, and beyond the DT at $z=10\,$pc. At all distances an equally well match of the lightcurves is possible. The (non-)availability of the BLR and DT photon fields as targets for pion production and as absorbers of \g-rays reveals itself in slightly different spectra at several tens of GeV. These should become explorable with the future Cherenkov Telescope Array. Interestingly, neutrinos could be an interesting discriminator for the emission region location, since the different levels of pion production at different distances from the black hole result in significantly different numbers of produced neutrinos. The calculated fluxes could be in reach of next-generation neutrino observatories.

The cloud parameters inferred from these three locations are listed in Tab.~\ref{tab:modelres}, and show that the different locations indeed have a significant influence. The cloud densities increase by 2 orders of magnitude from the location at the outer edge of the BLR to the location beyond the DT. The inferred densities are too small for being BLR clouds, but match well estimates of densities in cores of star forming regions. While the inferred densities and total masses also match those in atmospheres of RGs, the radii of the clouds are too large. However, if the interaction region would be placed even further away from the black hole, the cloud size would \change{become comparable to} the sizes of RG atmospheres.

As in paper I, we have inferred cloud temperatures much below the temperature of the CMB, which is implausible for such large structures. On the other hand, we have assumed that all material of the cloud is carried along with the jet \change{and accelerated into the non-thermal spectrum. However, a significant fraction of particles that enter the jet, could remain thermal, and an even larger fraction of the cloud material might not even be injected into the jet during the crossing. In such a case,} the original density and temperature of the cloud would be significantly higher.

Furthermore, we have tried to reproduce six of the days-long subflares in a fourth simulation. We used Example II as the baseline model. Assuming that the subflares originate from substructures in the larger cloud, which can be described similarly to the large cloud, we inferred the subcloud parameters. Their total masses are on the same order of magnitude as the large cloud implying a much larger density. They do, in fact, resemble densities of BLR clouds. Apart from the used location outside the BLR, the fact that the larger cloud does not fit BLR parameters, speaks against \change{the BLR scenario for the subclouds}. Size and density of the subclouds could resemble both RG atmospheres or the inner parts of star-forming regions. %These seem to be the more plausible scenarios for the entire 4-month outburst including the subflares.

The cloud originating from a star-forming region could be in line with the evolution of \cta\ over the last few years. Analyzing all available data from the used instruments since mid-2008, we have found that the source was active from late 2015 till early 2018. During that time it never returned to quiescence levels of previous years. Hence, the 4-month outburst discussed in detail could be part of a connected 2-year-long event. The idea that a molecular cloud, or a larger star-forming region, collides with the jet and the slow cut through the region by the jet creates years-long activity with flares of different duration and power, is certainly intriguing. Such a scenario could take place in other active galaxies as well, where long-lasting activity is observed. It is, however, difficult to test. Detection of molecular gas and dust in blazar host galaxies
% JPL: not "AGN", which is too general. Gas and dust are commonly detected e.g. in Seyfert galaxies. 
% MZ: fair point. On second thought, "dust" might be easy to be mistaken with the dusty torus. So, I added "host galaxies" to put the emphasize that I mean it really in the galaxy and not (yet) too close to the black hole. 
would be an asset. While this is obvious for the nearby radio galaxy Centaurus A, it is much more difficult for distant blazars, where the AGN outshines the host galaxy.

%
%################################################################################################################################
%
%ACKNOWLEDGMENTS
\acknowledgments
The authors wish to thank Patrick Kilian for stimulating discussions and significant help to improve the efficiency of the hadronic code. We also thank Frank Rieger for pointing out that only a fraction of particle is accelerated at a shock. Helpful comments and suggestions by the anonymous referee, which significantly improved the manuscript, are greatfully acknowledged.

The work of M.~Z. and M.~B. is supported through the South African Research Chair Initiative (SARChI) of the South African Department of Science and Technology (DST) and National Research Foundation.\footnote{Any opinion, finding and conclusion or recommendation expressed in this material is that of the authors, and the NRF does not accept any liability in this regard.}
F.~J. and S.J.~W. acknowledge support by the German Ministry for Education and Research (BMBF) through Verbundforschung Astroteilchenphysik grant 05A11VH2.
J.-P.~L. gratefully acknowledges CC-IN2P3 (\href{https://cc.in2p3.fr}{cc.in2p3.fr}) for providing a significant amount of the computing resources and services needed for this work.
A.~W. is supported by the Foundation for Polish Science (FNP).
%
%################################################################################################################################
%

%

%
%################################################################################################################################
%
\appendix
\section{Cloud ablation by the relativistic jet} \label{app:theory}
%
%\begin{figure*}[th]
%\begin{minipage}{0.49\linewidth}
%\centering \resizebox{\hsize}{!}
%{\includegraphics{clouddens.pdf}}
%\end{minipage}
%\hspace{\fill}
%\begin{minipage}{0.49\linewidth}
%\centering \resizebox{\hsize}{!}
%{\includegraphics{cloudnum.pdf}}
%\end{minipage}
%\caption{{\bf (a)} Numerical (solid) and approximate (dashed) solution for the cloud density distribution $n_c\p$, Eq. (\ref{eq:hydrostat2}), as a function of cloud radius $r_c\p$ for two values of the cloud temperature $T_c\p$ as labeled. The central density is set to $n_0\p = 10^{10}\,$cm$^{-3}$. {\bf (b)} Integration of the numerical solution (solid) and the analytical solution, Eq. (\ref{eq:ns}), (dashed) of the cloud density distribution as a function of slice position $x\p$ for two values of its temperature $T_c\p$ as labeled. Parameters as in (a). In both panels, we dropped the primes for clarity.}
%\label{fig:cloud}
%\end{figure*} 
%
%
\begin{figure}[th]
\centering
\includegraphics[width=0.25\textwidth]{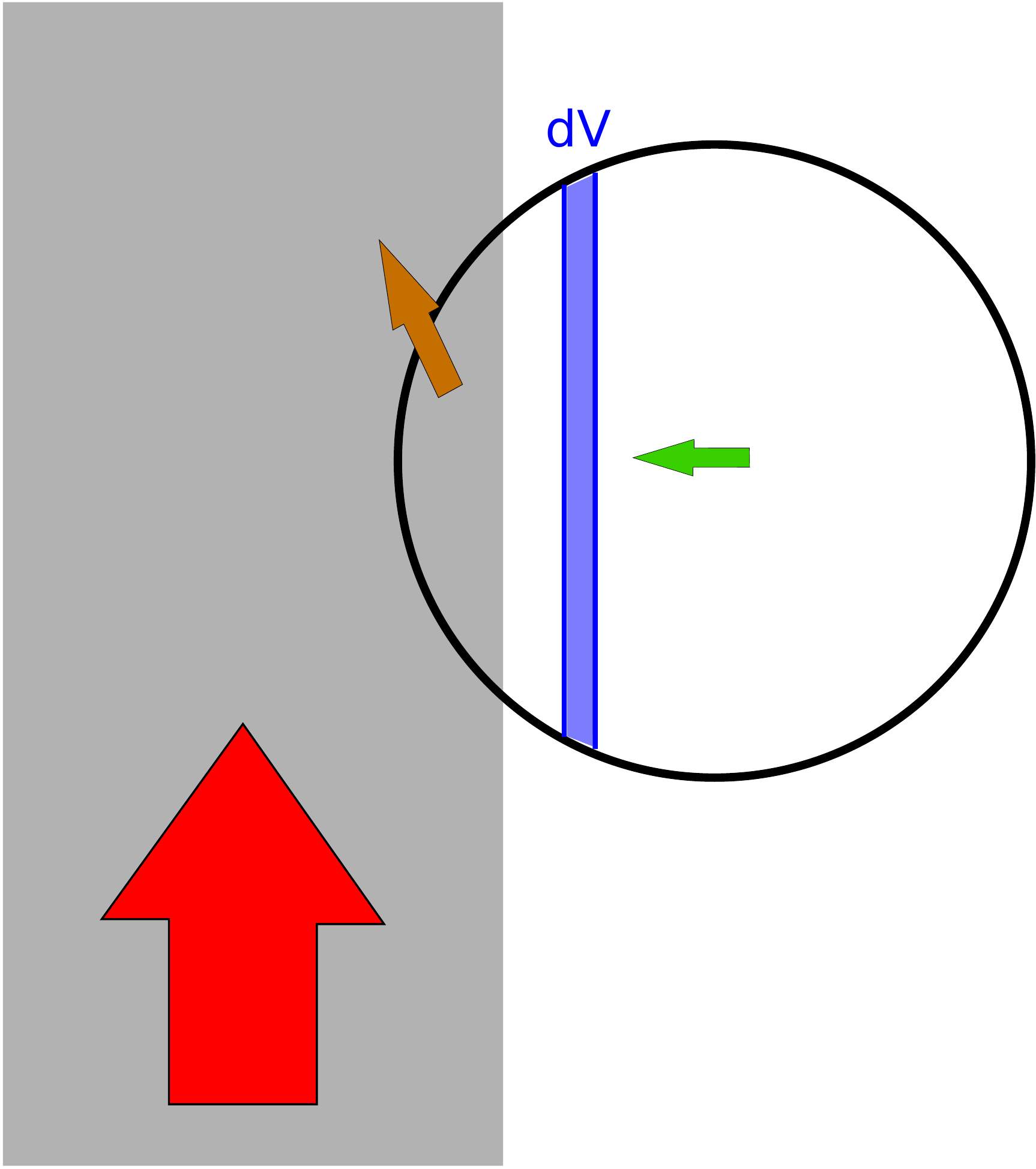}
\caption{Sketch of the ablation of a gas cloud. The gray area is a part of the jet with the red arrow depicting its ram pressure. The cloud is the black sphere with its velocity component orthogonal to the jet indicated by the green arrow. The brown arrow depicts the motion of the cloud material that has been ablated by the jet (parts of the cloud within the jet). The blue area is an example how the volume increment $\td{V}$ changes with time.}
\label{fig:cloud}
\end{figure} 
%
% SW: The green arrow could be misinterpreted as that clouds can ONLY enter orthogonally to the jet. One should add that the arrow represents the orthogonal component.
% MZ: Done.
Here, we briefly summarize the outline of the ablation model, which has been discussed in greater detail in paper I.

Consider a spherical cloud approaching the jet (cf., Fig. \ref{fig:cloud}) with orbital speed around the central black hole
\begin{align}
 v_c\p = \sqrt{G M_{\rm bh}/z\p} \label{eq:vc} ,
\end{align}
where $G$ is the gravitational constant, and $z\p$ the distance between the cloud and the black hole. The radius of the cloud can be derived from the rising time $t_f\p$ (that is from the beginning to the peak) of the outburst:
\begin{align}
 R_c\p = t_f\p v_c\p = \frac{t_f^{\rm obs} v_c\p}{(1+\zred)}  \label{eq:Rc} .
\end{align}
Apart from the redshift correction, the frame of the cloud and the observer's frame are identical, since the motion of the cloud is non-relativistic. Hence, the observed duration of the flare is indeed the same as the cloud penetration time.

The jet's ram pressure causes the immediate ablation of the cloud, if the gravitational pressure of the cloud cannot keep it together. As was shown in paper I, the cloud will hardly withstand ablation even if the jet's protons are cold in the comoving frame. Here, we also consider relativistic protons, which increase the jet's ram pressure substantially. The cloud's destruction is even more likely in this scenario.

Given that the cloud penetrates the jet gradually, the number of particles injected into the jet changes over time, which is indicated in Fig. \ref{fig:cloud}. Both the density profile and the variable volume element that is ablated, must be considered.

%In order to calculate the correct injection term, the density distribution of the cloud $n_c\p(r_c\p)$ must be known. 
We consider again a profile based on hydrostatic equilibrium, and neglect additional pressure constituents, such as turbulence, inhomogeneities and the magnetic field of the cloud. Hence, the cloud consists of isothermal ideal gas with temperature $T_c\p$, so that the thermal pressure $p_T\p = \rho_c\p k_B T_c\p / m_p$, where $\rho_c\p = m_p n_c\p$ is the cloud's mass density, and $k_B$ is the Boltzmann constant. The equation of hydrostatic equilibrium then reads
\begin{align}
 \frac{k_B T_c\p}{m_p} \frac{\td{\rho_c\p(r_c\p)}}{\td{r_c\p}} &= - g(r_c\p) \, \rho_c\p(r_c\p) \nonumber \\
 &= - 4 \pi \frac{G \rho_c\p(r_c\p)}{r_c^{\prime 2}} \int\limits_{0}^{r_c\p} \td{\tilde{r}} \tilde{r}^2 \rho_c\p(\tilde{r}) \label{eq:hydrostat} .
\end{align}
With the definition $\tau\p \equiv k_B T_c\p / (4 \pi \, m_p \, G)$, Eq. (\ref{eq:hydrostat}) reduces to
\begin{align}
 \tau\p \, \frac{\td{}}{\td{r_c\p}} \left( \frac{r_c^{\prime 2}}{\rho\p} \, \frac{\td{\rho\p}}{\td{r_c\p}} \right) = - \rho\p \, r_c^{\prime 2} \label{eq:hydrostat2} .
\end{align}
As we have found in paper I, the solution to this differential equation is well approximated by:
\begin{align}
 n_c\p(r_c\p) = \frac{n_0\p}{1+\left( r_c\p / r_0\p \right)^2} \label{eq:ncloud} .
\end{align}
The normalization $n_0\p$ can be determined by integrating Eq. (\ref{eq:ncloud}) and equating it to the total number of particles in the cloud. The scale height is defined as 
\begin{align}
 r_0\p = \sqrt{\frac{3\tau\p}{m_p n_0\p}} \label{eq:r0p} .  
\end{align}

We again approximate the cloud as a sphere with outer boundary $R_c\p>r_0\p$ and set $n_c\p(r_c\p\geq R_c\p)=0$. As depicted in Fig. \ref{fig:cloud}, the cloud is ablated slice-by-slice. Therefore, we define all quantities of the cloud as a function of $x\p$, the slice position with respect to the outer edge of the cloud that first touches the jet. That is, $x\p=0$ where the cloud first touches the jet, $x\p=R_c\p$ is the cloud's center, and $x\p=2R_c\p$ marks the rear side of the cloud. With the speed of the cloud, it can be written as $x\p = v_c\p t\p$, where $t\p$ is the time that has passed since first contact in the AGN frame.

The number of particles ablated in each slice is the integral over the density $n_c\p(r_c\p)$ with respect to the slice volume $\td{V_s}$. In the case of a sphere, the volume of a slice between positions $x\p$ and $x\p+\td{x\p}$ is \citep{zs13}
\begin{align}
 \td{V_s\p(x\p)} = \td{x\p} \int \td{A_s\p(x\p)} = \pi \left( 2R_c\p x\p-x^{\prime 2} \right) \td{x\p} \label{eq:Vslice} ,
\end{align}
where $A_s\p(x)$ is the cross-section of a slice, and $\td{x\p}$ its width.
The particle number in each slice then becomes
%\begin{align}
% \td{N_s\p(x\p)} = \td{x\p} \int n_c\p(r_c\p) \td{A_s\p(x\p)} \label{eq:ns1} .
%\end{align}
%Writing the integral in cylindrical coordinates 
with $r_c\p(x\p) = \sqrt{\omega^2 + (R_c\p-x\p)^2}$, and $\omega_c\p(x\p) = \sqrt{2R_c\p x\p-x^{\prime 2}}$:
%, Eq. (\ref{eq:ns1}) becomes
\begin{align}
 \td{N_s\p(x\p)} = 2\pi\td{x\p} \intl_0^{\omega_c\p(x\p)} n_c\p(r_c\p(\omega))\, \omega \td{\omega} \label{eq:ns2} .
\end{align}
Inserting Eq. (\ref{eq:ncloud}) in Eq. (\ref{eq:ns2}), we find
\begin{align}
 \td{N_s\p(x\p)} = \pi\td{x\p}r_0^{\prime 2}n_0\p \logb{\frac{r_0^{\prime 2} + R_c^{\prime 2}}{r_0^{\prime 2}+(R_c\p-x\p)^2}} \label{eq:ns} .
\end{align}
%This function is shown in Fig. \ref{fig:cloud}(b) for two cases of $T_c$ along with an integration of the numerical solution of Eq. (\ref{eq:hydrostat2}). The analytical approximation and the exact result match nicely.

The injection of particles in the jet %, which get dragged along and cause the flare at a shock somewhere downstream, 
can then be described by
\begin{align}
 Q_{\rm inj}(t) \propto \logb{\frac{r_0^{\prime 2} + R_c^{\prime 2}}{r_0^{\prime 2}+(R_c\p-x\p)^2}} \DF{t-\frac{x\p}{v_c\p}} \label{eq:injection} .
\end{align}
Here, $\DF{q}$ is Dirac's $\delta$-function, which describes the slice-by-slice ablation in time.

%We stress that the entire mass of the cloud is not added to the jet at once, but gradually over about 4 months in the observer's frame. Hence, the impact of the added mass on the jet's bulk Lorentz factor at any given time is minor compared to a case where the entire cloud mass would be added at once. In the following, we assume a constant jet bulk Lorentz factor.

From the modeling of the jet emission, we can infer the cloud parameters. Since the flux variation is mainly due to changes in the number of injected particles, the difference of particle number $\Delta N$ between the number of injected particles during the peak of the flare and the beginning of the flare provides the (minimum) number of particles in the cloud. The cloud's density then simply becomes
\begin{align}
 n\p_c = \frac{\Delta N}{\frac{4}{3}\pi R_c^{\prime 3}} \label{eq:clouddens}.
\end{align}
Since the scale height of the cloud, Eq. (\ref{eq:r0p}), depends on the temperature of the gas, the latter can be calculated. As shown in paper I, the temperature is
\begin{align}
 T\p_c \simeq 0.4 \frac{G m_p^2 \Delta N}{k_B R\p_c} \label{eq:cloudtemp}.
\end{align}

%
%################################################################################################################################
%
\section{The hadronic emission model} \label{app:hadmod}
\begin{figure*}[th]
\centering
\includegraphics[width=0.75\textwidth]{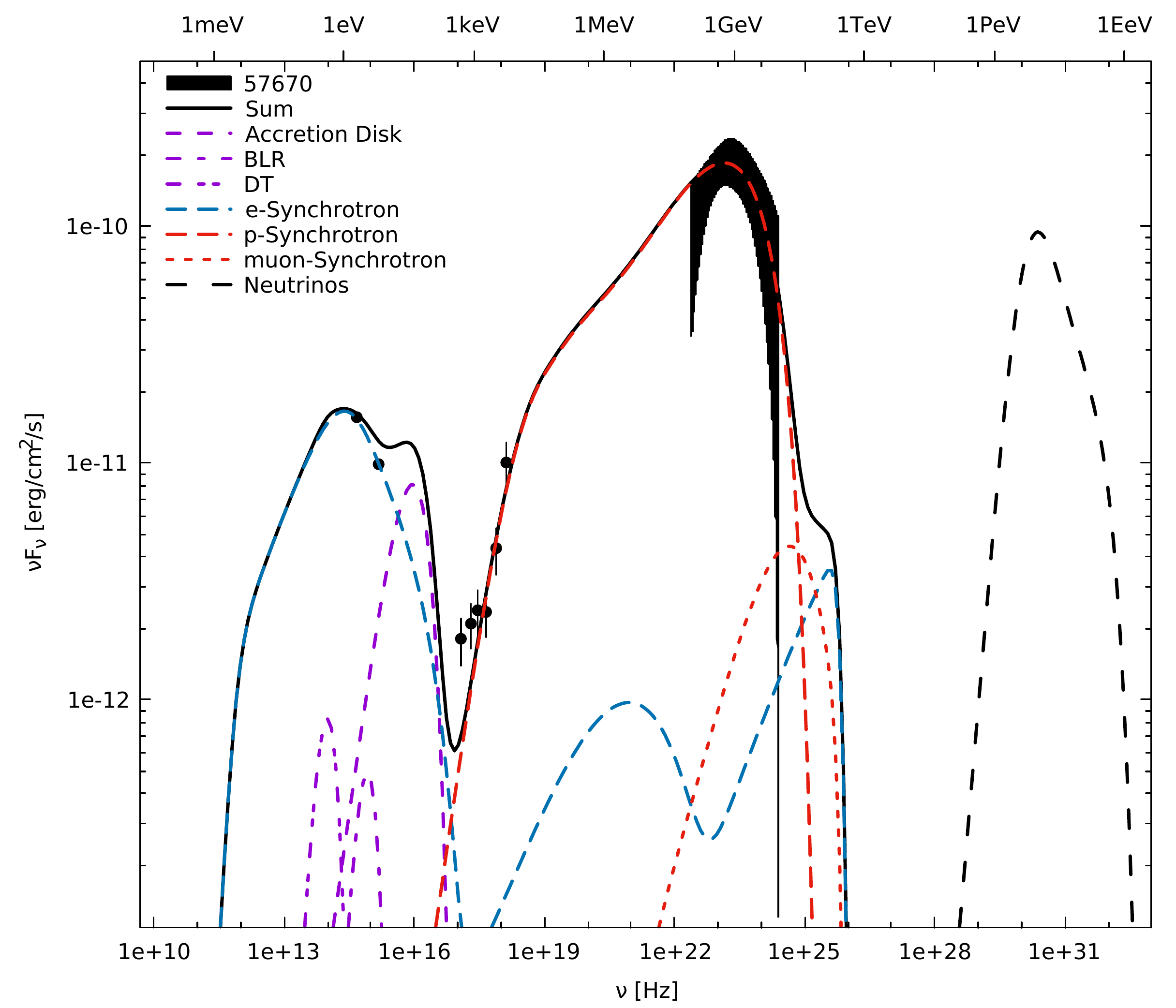}
\caption{SED with all emission processes that have a visible impact on the total spectrum (see legend). As an example we have chosen the data of MJD 57670 and the corresponding spectral model from Example II.}
\label{fig:det-spec}
\end{figure*} 
The numerical code used here to describe the hadronic model is based upon \cite{dbf15}. It is a one-zone model, where a spherical region of radius $R$, and with tangled magnetic field $B$ produces the entire radiative output. Using a Crank-Nicholson scheme, the code solves for each particle species (protons, pions, muons, electrons/positrons) a Fokker-Planck-type equation including terms for injection, stochastic acceleration, continuous losses (radiative and adiabatic cooling), and catastrophic losses (escape and particle decay). The escape is treated as energy-independent and as a multiple $\eta_{\rm esc}$ of the light-crossing time scale $R/c$. The acceleration term is also energy independent and, in turn, parameterized as a multiple $\eta_{\rm acc}$ of the escape time scale. The escape and acceleration term are equal for all particle species. The acceleration term is considered as a mild re-acceleration of particles through the present magnetic field, but it is not considered as the primary acceleration of the particles to their initial high energies. The primary acceleration could be accounted for by a separated acceleration zone as in the models of \cite{ws15} and \cite{cpb15}, but is not explicitly modeled here.

The primary particles (protons and electrons) are injected with single power-laws with minimum Lorentz factor $\gamma_{\rm min}^{i}$, maximum Lorentz factor $\gamma_{\rm max}^{i}$, spectral index $s^{i}$, and injection luminosity $L_{\rm inj}^{i}$ (the superscript $i$ is either protons (+) or electrons (-)). Protons are subject to synchrotron and adiabatic losses, as well as to pion production through collisions with ambient soft photon fields. The charged pions are subject to synchrotron cooling, even though their fast decay into muons and neutrinos does not allow them to produce a significant amount of synchrotron radiation. The muons, on the other hand, exist long enough to produce a significant amount of synchrotron emission before decaying into electrons/positrons and neutrinos. The thusly created electrons and positrons are used as an additional injection on top of the primary electrons. They are subject to synchrotron losses and inverse Compton losses on ambient soft photon fields. For all particle species the possible radiation spectra are calculated, as are the integrated fluxes in given energy bands and the neutrino output. An example spectrum with all contributions is shown in Fig.~\ref{fig:det-spec}.

To introduce the time-dependency, after finding an initial equilibrium for all particle species, parameters of choice can be changed. At each time step, the previously calculated particle distributions are kept and another set of primary particles is injected on top. In the end, one obtains spectra for each time step, as well as light curves in given energy bands.

We have modified the code to accommodate for external soft photon fields, such as the accretion disk, the BLR, and the dusty torus. In order to calculate the external absorption, we included the description of \cite{be16}. Furthermore, the external fields are used as target photons for pion production by protons, and inverse Compton emission by electrons. These processes, of course, depend on the distance $z\p$ of the emission region from the black hole.

\change{
%###############################################################################################################################
\section{Example IV: Modeling the subflares} \label{sec:modIV}
}
\begin{table*}[th]
\caption{Variability parameters used for the subflares. Parameter description, symbol and value. The MJD peaks give the time of maximum flux in the \g-ray and optical band for the subflare, respectively. The number in brackets give the number of days $m$ for the variability.}
\begin{tabular}{lcl|c|c|c|c|c|c}
Definition	& \multicolumn{2}{c|}{Symbol} 	& IV.1	& IV.2	& IV.3	& IV.4	& IV.5	& IV.6  \\
\hline
MJD peak (\g-rays)	& $\mathscr{T}_{\rm \g,sf}$	& 	& $57715$	& $57724$	& $57738$	& $57744$	& $57750$	& $57759$ \\
MJD peak (optical)	& $\mathscr{T}_{\rm opt,sf}$	& 	& $57716$	& $57724$	& $57738$	& $57745$	& $57752$	& $57759$ \\
Proton injection variability	& $\mathscr{L}_{\rm p,sf}$	& [erg/s]	& $7.0\times 10^{44}$ (2)	& $2.0\times 10^{44}$ (1)	& $6.5\times 10^{44}$ (2)	& $1.1\times 10^{45}$ (0)	& $2.0\times 10^{45}$ (0)	& $9.0\times 10^{44}$ (3) \\
Proton max. Lorentz factor variability	& $\mathscr{\gamma}_{\rm p,max,sf}$	& 			& $-0.3\E{9}$ (2) 	& $0.3\E{9}$ (2)	& $0.2\E{9}$ (2) 	& $-0.3\E{9}$ (0)	& $0.1\E{9}$ (1)	& - \\
Electron injection variability	& $\mathscr{L}_{\rm e,sf}$	& [erg/s]	& $1.2\E{42}$ (3)	& $3.2\times 10^{42}$ (2)	& $5.6\times 10^{42}$ (2) 	& $6.4\times 10^{42}$ (0)	& $9.6\times 10^{42}$ (2)	& $6.4\times 10^{42}$ (3) \\
Magnetic field variability	& $\mathscr{B}_{\rm sf}$	& [G]	& $-10$ (2)	& $10$ (2)	& $20$ (2) 	& $10$ (0)	& $10$ (1)	& $10$ (3)
\end{tabular}
\label{tab:modelparamIV}
\end{table*}
\begin{table*}[th]
\caption{Summary of the model results for the subflares. The minimum gives the value at the beginning and the maximum at the peak of the subflare. The Eddington power of CTA\,102's black hole is $P\obs_{\rm Edd}=1.1\E{47}\,$erg/s. Parameters below the horizontal line give the inferred sub-cloud parameters.}
\begin{tabular}{lcl|c|c|c|c|c|c}
Definition		& \multicolumn{2}{c|}{Symbol} 						& IV.1				& IV.2				& IV.3	& IV.4				& IV.5				& IV.6  \\
\hline
Minimum proton power		& $P\obs_{\rm p,min}$	& [erg/s]		& $6.1\E{47}$	& $8.1\E{47}$	& $8.7\E{47}$	& $1.2\E{48}$	& $1.2\E{48}$	& $1.2\E{48}$ \\
Maximum proton power		& $P\obs_{\rm p,max}$	& [erg/s]		& $2.5\E{48}$	& $1.2\E{48}$	& $2.4\E{48}$	& $2.5\E{48}$	& $3.6\E{48}$	& $3.4\E{48}$ \\
Minimum electron power		& $P\obs_{\rm e,min}$	& [erg/s]		& $5.1\E{42}$	& $7.2\E{42}$	& $1.2\E{43}$	& $1.6\E{43}$	& $1.7\E{43}$	& $1.5\E{43}$ \\
Maximum electron power		& $P\obs_{\rm e,max}$	& [erg/s]		& $1.7\E{43}$	& $1.9\E{43}$	& $3.0\E{43}$	& $4.7\E{43}$	& $4.9\E{43}$	& $3.9\E{43}$ \\
Minimum radiative power		& $P\obs_{\rm r,min}$	& [erg/s]		& $1.4\E{46}$	& $1.9\E{46}$	& $2.7\E{46}$	& $3.9\E{46}$	& $4.4\E{46}$	& $3.7\E{46}$ \\
Maximum radiative power		& $P\obs_{\rm r,max}$	& [erg/s]		& $4.1\E{46}$	& $3.9\E{46}$	& $1.3\E{47}$	& $9.5\E{46}$	& $1.6\E{47}$	& $1.3\E{47}$ \\
Minimum magnetic power		& $P\obs_{\rm B,min}$	& [erg/s]		& $6.6\E{48}$	& $6.6\E{48}$	& $6.6\E{48}$	& $6.6\E{48}$	& $6.6\E{48}$	& $6.6\E{48}$ \\
Maximum magnetic power		& $P\obs_{\rm B,max}$	& [erg/s]		& $4.6\E{48}$	& $9.0\E{48}$	& $1.2\E{49}$	& $9.0\E{48}$	& $9.0\E{48}$	& $9.0\E{48}$ \\
Minimum proton density		& $n_{\rm p,min}$		& [cm$^{-3}$]	& $4.0\E{-3}$	& $5.7\E{-3}$	& $4.8\E{-3}$	& $6.7\E{-3}$	& $6.7\E{-3}$	& $6.8\E{-3}$ \\
Maximum proton density		& $n_{\rm p,max}$		& [cm$^{-3}$]	& $1.4\E{-2}$	& $6.9\E{-3}$	& $1.2\E{-2}$	& $1.1\E{-2}$	& $1.5\E{-2}$	& $1.7\E{-2}$ \\
Minimum electron density	& $n_{\rm e,min}$		& [cm$^{-3}$]	& $2.3\E{1}$	& $3.3\E{1}$	& $5.4\E{1}$	& $7.5\E{1}$	& $8.2\E{1}$	& $6.9\E{1}$ \\
Maximum electron density	& $n_{\rm e,max}$		& [cm$^{-3}$]	& $7.7\E{1}$	& $8.3\E{1}$	& $1.3\E{2}$	& $2.1\E{2}$	& $2.4\E{2}$	& $1.7\E{2}$ \\
\hline
Cloud speed							& $v_{\rm c}$	& [cm/s]		& $1.9\E{8}$	& $1.9\E{8}$	& $1.9\E{8}$	& $1.9\E{8}$	& $1.9\E{8}$	& $1.9\E{8}$ \\ 
Cloud radius						& $R_{\rm c}$	& [cm]			& $3.3\E{13}$	& $2.4\E{13}$	& $2.4\E{13}$	& $8.1\E{12}$	& $2.4\E{13}$	& $3.3\E{13}$ \\ 
Cloud density						& $n_{\rm c}$	& [cm$^{-3}$]	& $2.7\E{10}$	& $6.0\E{10}$	& $8.8\E{10}$	& $4.4\E{12}$	& $1.8\E{11}$	& $5.2\E{10}$ \\ 
Cloud mass							& $M_{\rm c}$	& [g]			& $6.8\E{27}$	& $5.8\E{27}$	& $8.5\E{27}$	& $1.6\E{28}$	& $1.7\E{28}$	& $1.3\E{28}$ \\ 
Cloud temperature					& $T_{\rm c}$	& [K]			& $3.0\E{-2}$	& $3.8\E{-2}$	& $5.5\E{-2}$	& $3.1\E{-1}$	& $1.2\E{-1}$	& $5.8\E{-2}$ 
\end{tabular}
\label{tab:modelresIV}
\end{table*}
Here, we extend the long-term variability model of section \ref{sec:res} to account for medium-term subflares. We fit the 4 bright subflares from mid-December to mid-January (for the first 3 we have contemporaneous spectra available) and 2 of the rising part of the long-term event, for which we also have contemporaneous spectra. With the available dense data for these six subflares, we can perform explicit modelings and analyze the differences in their parameter sets. The peak dates for the subflares are given in Tab.~\ref{tab:modelparamIV}. As the baseline, we use Example II for which the emission region is placed at a distance of $1\,$pc from the black hole -- within the DT. This choice is arbitrary. However, the inferred parameters for the underlying clouds could be scaled to the other distances using the results in Tab.~\ref{tab:modelres}. The subflares are modeled by varying the proton and electron injection power, as well as the maximum proton Lorentz factor and the magnetic field. For the latter two we use the constraints from Tab.~\ref{tab:paramconst}.

The variability of the subflares is modeled as follows. For the proton and electron injection we use a step-function injection:
\begin{align}
 L_{\rm i,inj}(t) = L_{\rm i,base}(t) + \mathscr{L}_{\rm i,sf} \SF{t}{\mathscr{T}_{\rm i,sf}-m}{\mathscr{T}_{\rm i,sf}} \label{eq:sfparinj},
\end{align}
where $L_{\rm i,base}(t)$ represents the baseline injection for particle species $i$ as in Eq.~(\ref{eq:cloudinj}), $\mathscr{L}_{\rm i,sf}$ is the subflare injection variation as listed in Tab.~\ref{tab:modelparamIV}, $\mathscr{T}_{\rm i,sf}$ is the peak time of the respective subflare (for protons we use the peak in the \g-ray light curve, and for electrons we use the peak in the optical light curves), $m$ is the number of days the subflare rises from the baseline to the peak (the number in brackets next to the variability value in Tab.~\ref{tab:modelparamIV}), and the step function is defined as
\begin{align}
 \SF{t}{t_1}{t_2} = 
 \begin{cases}
  1 &\mbox{if } t_1\leq t\leq t_2 \\
  0 &\mbox{else}
 \end{cases} \label{eq:stepfunction}.
\end{align}
Note that the step function includes both borders. Hence, for $m=2$ the injection for the subflare lasts 3 days.

The variability of the proton maximum Lorentz factor and the magnetic field is modeled differently. In these cases we assume a linear in- and decrease of the form
\begin{align}
 \gamma_{\rm p,max}(t) &= \gamma_{\rm p,max} + \mathscr{\gamma}_{\rm p,max,sf} \quad  g(t,\mathscr{T}_{\rm \g,sf},m) \nonumber \\
 &\quad\times \SF{t}{\mathscr{T}_{\rm \g,sf}-m}{\mathscr{T}_{\rm \g,sf}+m} \label{eq:sfgamvar} \\
 B(t) &= B + \mathscr{B}_{\rm sf} \quad g(t,\mathscr{T}_{\rm \g,sf},m) \nonumber \\
 &\quad\times \SF{t}{\mathscr{T}_{\rm \g,sf}-m}{\mathscr{T}_{\rm \g,sf}+m} \label{eq:sfBvar}
\end{align}
with
\begin{align}
 g(t,\mathscr{T}_{\rm \g,sf},m) = \frac{(m+1)-|(m+1)-(t-\mathscr{T}_{\rm \g,sf}+(m+1))|}{m+1} \label{eq:sflinvar}.
\end{align}
For $t=\mathscr{T}_{\rm \g,sf}\pm(m+1)$ the function $g=0$, while for $t=\mathscr{T}_{\rm \g,sf}$ the function $g=1$ with a linear in- and decrease in between. Note that both the $\gamma_{\rm p,max}$- and $B$-variability are centered on the peak of the \g-ray light curve, since the constraints on both parameters are connected through Eq.~(\ref{eq:FBg}).

The choices for the time dependency of the parameters are arbitrary in this case, however different choices would only have a minor influence on the outcome \change{given the short duration of the events}. The linear decrease in the magnetic field after the end of the particle injection results in a larger synchrotron loss of the particles during the decreasing part of the light curve.

The peak times of the \g-ray and optical light curve given in Tab.~\ref{tab:modelparamIV} indicate that the light curves are not fully correlated, since for three out of the six subflares the optical light curve peaks after the \g-ray light curve. While this might be a sampling effect, we can account for this by delaying the optical injection with respect to the proton injection. While this can produce a nice fit, it is difficult to explain why the electrons should be injected/accelerated after the protons. Fig.~\ref{fig:mod-lc} also shows that the X-ray light curve is not fully correlated with the \g-ray light curve with either delayed peaks or none at all. This is not possible to reproduce with our model. The (mostly) slow cooling of the protons implies that the light curve is dominated through the injection and not through cooling effects, which could explain a delayed response of the X-ray light curve.

With the variability parameters of Tab.~\ref{tab:modelparamIV} and Eqs.~(\ref{eq:sfparinj}) to (\ref{eq:sfBvar}), we model the six subflares. The model is shown as the dash-dotted lines in Fig.~\ref{fig:mod-lc} for the light curves and in Fig.~\ref{fig:mod-spec} for the spectra. Apart from the already mentioned caveats, the data are well fit, especially the spectra during the subflares are well represented. Fig.~\ref{fig:mod-part-I}(IV) shows the particle distributions and time scales. Especially the changes in the proton distribution through the variations in the maximum Lorentz factor are evident. The changes in the magnetic field result in changing cooling time scales for both protons and electrons. The influence on the particle distributions by the variable cooling time scales is, however, minor.

Tab.~\ref{tab:modelresIV} lists the minimum and maximum powers for each subflare, as well as the proton and electron densities. Note that the "maximum" should be interpreted as the time of the maximum in the light curve. This mostly coincides also with the maximum power output, unless the magnetic field is decreased during a subflare. As for the long-term Examples I to III, the power is dominated by the magnetic field, followed by the proton power. The radiative and electron powers are minor. This even holds for the first subflare (IV.1), for which the magnetic field is decreased. 

However, in order to account for the subflares in the different bands, the particle injections actually behave differently for each flare. As already mentioned, the peaks of the subflares are not always at the time in different energy bands. Secondly, the ratio of injected protons and electrons is also changing from flare to flare. Subflares IV.1, IV.3, and IV.6 require a higher relative proton injection compared to the electrons, while it is the opposite for the three other subflares. This reflects the different relative flux changes between the \g-ray domain and the optical bands. While the independence of the particle distributions in the hadronic model allows for an easier representation of these changes than a leptonic one-zone model, it remains unclear why the similar conditions during the subflare should lead to distinctively different behaviors and outcomes -- e.g., why the relative number of accelerated particles should change, while the magnetic field is similar.

Nevertheless, we can use the model parameters to estimate the cloud parameters, if we assume that the subflares are caused by substructures in the larger cloud and that these substructures are governed by roughly the same equations. The resulting parameters are listed in Tab.~\ref{tab:modelresIV}. Naturally, the speed of each small cloud is the same as the larger cloud, from which we can calculate the radii. Since these small clouds are responsible for very strong outbursts, their densities are much higher than that of the larger cloud, and in turn the temperatures are also higher -- however, also not at a realistic value \change{implying that the arguments of section \ref{sec:cloud} hold here, as well.}


\begin{thebibliography}{}

\bibitem[Aarsten et al. (2018)]{aarst18}
Aarsten, M.~G., et al. (The IceCube Collaboration, {\it Fermi}-LAT, MAGIC, {\it AGILE}, ASAS-SN, HAWC, H.E.S.S., {\it INTEGRAL}, Kanata, Kiso, Kapteyn, Liverpool Telescope, Subaru, {\it Swift}/{\it NuSTAR}, VERITAS, VLA/17B-403 teams), 2018, Science, 6398, 147

\bibitem[Acharya et al.(2013)]{acha13}
Acharya, B.~S., Actis, M., Aghajani, T., et al., 2013, Astroparticle Physics, 43, 3 

\bibitem[Acero et al.(2015)]{aFea15}
Acero F., Ackermann M., Ajello M., et al., 2015, ApJS, 218, 23

\bibitem[Acero et al.(2016)]{aFea16}
Acero F., Ackermann M., Ajello M., et al., 2016, ApJS, 223, 26

%\bibitem[Araudo et al.(2009)]{abr09}
%Araudo A.T., Bosch-Ramon V., Romero G.E., 2009, A\&A, 503, 673

%\bibitem[Araudo et al.(2010)]{abr10}
%Araudo A.T., Bosch-Ramon V., Romero G.E., 2010, A\&A, 522, A97

\bibitem[Arnaud(1996)]{a96}
Arnaud K.A., 1996, ASPC, 101, 17

\bibitem[Atwood et al.(2009)]{aFea09}
Atwood W.B., Abdo A.A., Ackermann M., et al., 2009, ApJ, 697, 1071

\bibitem[Bachev et al.(2017)]{bea17}
Bachev R., Popov V., Strigachev A., Semkov E., et al., 2017, MNRAS, 471, 2216

\bibitem[Barkov et al.(2010)]{bab10}
Barkov M.V., Aharonian F.A., Bosch-Ramon V., 2010, ApJ, 724, 1517

%\bibitem[Blandford \& K\"onigl(1979)]{bk79}
%Blandford R., K\"onigl A., 1979, ApL 20, 15

\bibitem[Blandford \& Rees(1974)]{br74}
Blandford R., Rees M.J., 1974, MNRAS, 169, 395

\bibitem[Blandford \& Znajek(1977)]{bz77}
\change{Blandford R., Znajek R.L., 1977, MNRAS, 179, 433}

%\bibitem[Bosch-Ramon(2015)]{b15}
%Bosch-Ramon V., 2015, A\&A, 575, A109

\bibitem[Bosch-Ramon et al.(2012)]{bpb12}
Bosch-Ramon V., Perucho M., Barkov M.V., 2012, A\&A, 539, A69

\bibitem[B\"ottcher \& Els(2016)]{be16}
B\"ottcher M., Els P., 2016, ApJ, 821, 102

\bibitem[Burrows et al.(2005)]{bea05}
Burrows D.N., Hill J.E., Nousek J.A., Kennea J.A., 2005, SSRv, 120, 165

\bibitem[Carroll \& Ostlie(2017)]{co17}
Carroll B.W., Ostlie D.A., 2017, ``An Introduction to Modern Astrophysics''. 2nd Edition, Cambridge University Press, DOI: 10.1017/9781108380980

\bibitem[Chen et al.(2015)]{cpb15}
Chen X., Pohl M., B\"ottcher M., 2015, MNRAS, 447, 530

\bibitem[Costamante et al.(2018)]{cea18}
Costamante L., Cutini S., Tosti G., Antolini E., Tramacere A., 2018, MNRAS, 477, 4749

\bibitem[de la Cita et al.(2017)]{dcea17}
de la Cita V.M., del Palacio S., Bosch-Ramnon V., Paredes-Fortuny X., Romero G.E., Khangulyan D., 2017, A\&A, 604, A39

\bibitem[Dietrich et al.(1999)]{dea99}
Dietrich M., Wagner S.J., Courvoisier T.J.-L., Bock H., North P., 1999, A\&A, 351, 31

\bibitem[Diltz et al.(2015)]{dbf15}
Diltz C., B\"ottcher M., Fossati G., 2016, ApJ, 802, 133

\bibitem[Franceschini et al.(2008)]{frv08}
Franceschini A., Rodighiero G., Vaccari M., 2008, A\&A, 487, 837

%\bibitem[Fromm et al.(2011)]{fea11}
%Fromm C.M., Perucho M., Ros E., Savolainen T., et al., 2011, A\&A, 531, A95

\bibitem[Gehrels et al.(2004)]{gea04}
Gehrels N., Chincarini G., Giommi P., Mason K.O., et al., 2004, ApJ, 611, 1005

%\bibitem[Ghisellini et al.(1998)]{gea98}
%Ghisellini G., Celotti A., Fossati G., Maraschi L., Comastri A., 1998, MNRAS, 301, 451

\bibitem[Giannios(2013)]{g13}
Giannios D., 2013, MNRAS, 431, 355

\bibitem[Giommi et al.(2006)]{gea06}
Giommi P., Blustin A.J., Capalbi M., Colafrancesco S., et al., 2006, A\&A, 456, 911

\bibitem[Hauser et al.(2004)]{hea04}
Hauser M., M\"ollenhoff C., P\"uhlhofer G., Wagner S.J., Hagen H.-J., Knoll M.,  2004, AN, 325, 659

\bibitem[Hayashida et al.(2012)]{Hea12}
Hayashida M., Madejski G.M., Nalewajko K., Sikora M., et al., 2012, ApJ, 754, 114

\bibitem[IceCube Collaboration(2013)]{icc13}
IceCube Collaboration, 2013, Science, 342, 1242856

\bibitem[Kalberla et al.(2005)]{kea05}
Kalberla P.M.W., Burton W.B., Hartmann D., Arnal E.M., et al., 2005, A\&A, 440, 775

%\bibitem[Klein et al.(1994)]{kkc94}
%Klein R.I., McKee C.F., Colella P., 1994, ApJ, 420, 213

%\bibitem[Komissarov(1994)]{k94}
%Komissarov S.S., 1994, MNRAS, 269, 394

\bibitem[Larionov et al.(2016)]{lea16}
Larionov V.M., Villata M., Raiteri C.M., Jorstad S.G., et al., 2016, MNRAS, 461, 3047

\bibitem[Lister et al.(2016)]{leaM16}
Lister M.L., Aller M.F., Aller H.D., Homan D.C., et al., 2016, AJ, 152, 12

\bibitem[Malmrose et al.(2011)]{mea11}
Malmrose M.P., Marscher A.P., Jorstad S.G., Nikutta R., Elitzur M., 2011, ApJ, 732, 116

\bibitem[Marscher(2014)]{m14}
Marscher A.P., 2014, ApJ, 780, 87

%\bibitem[Mattox et al.(1996)]{mea96}
%Mattox J.R., Bertsch D.L., Chiang J., et al., 1996, ApJ, 461, 396

%\bibitem[Perucho et al.(2017)]{pbb17}
%Perucho M., Bosch-Ramon V., Barkov M.V., 2017, A\&A, 606, A40

\bibitem[Perucho et al.(2014)]{pmlh14}
Perucho M., Marti J.M., Laing R.A., Hardee P.E., 2014, MNRAS, 441, 1488

\bibitem[Peterson(2006)]{p06}
Peterson B.M., 2006, LNP, 693, 77

\bibitem[Petropoulou et al.(2017)]{pea17}
Petropoulou M., Nalewajko K., Hayashida M., Mastichiadis A., 2017, MNRAS, 467, L16

\bibitem[Pian et al.(2005)]{pft05}
Pian E., Falomo R., Treves A., 2005, MNRAS, 361, 919

%\bibitem[Poludnenko et al.(2002)]{pfb02}
%Poludnenko A.Y., Frank A., Blackman E.G., 2002, ApJ, 576, 832

\bibitem[Raiteri et al.(2017)]{rea17}
Raiteri C.M., Villata M., Acosta-Pulido J.A., Agudo I., et al., 2017, Nature, 552, 374

\bibitem[Romoli et al.(2018)]{Rea18}
Romoli C., Zacharias M., Meyer M., Ait Benkhali F., Jacholkowska A., Wierzcholska A., Jankowsky F., Lenain J.-P., 2018, PoS, 35, 649

\bibitem[Roustazadeh \& B\"ottcher(2012)]{rb12}
Roustazadeh P., B\"ottcher M., 2012, ApJ, 750, 26

\bibitem[Sahai et al.(2013)]{sea13}
\change{Sahai R., Vlemmings W.H.T., Huggins P.J., Nyman L.-\AA., Gonidakis I., 2013, ApJ, 777, 92}

\bibitem[Schlafly \& Finkbeiner(2011)]{sf11}
Schlafly E.F., Finkbeiner D.P., 2011, ApJ, 737, 103

%\bibitem[Schramm et al.(1993)]{sea93}
%Schramm K.-J., Borgeest U., Camenzind M., Wagner S.J., et al., 1993, A\&A, 278, 391

%\bibitem[Shakura \& Sunyaev(1973)]{ss73}
%Shakura N.I., Sunyaev R.A., 1973, A\&A, 24, 337

\bibitem[Shukla et al.(2018)]{sea18}
Shukla A., Mannheim K., Patel S.R., Roy J., et al., 2018, ApJL, 854, L26

\bibitem[Weidinger \& Spanier(2015)]{ws15}
Weidinger M., Spanier F., 2015, A\&A, 573, A7

\bibitem[Zacharias \& Schlickeiser(2013)]{zs13}
Zacharias M., Schlickeiser R., 2013, ApJ, 777, 109

\bibitem[Zacharias et al.(2017)]{zea17}
Zacharias M., B\"ottcher M., Jankowsky F., Lenain J.-P., Wagner S., Wierzcholska A., 2017, ApJ, 851, 72 {\bf (paper I)}

\bibitem[Zacharias et al.(2018)]{Zea18}
Zacharias M., Sitarek J., Dominis Prester D., Jankowsky F., Lindfors E., Mohamed M., Sanchez D., Terzic T., 2018, PoS, 35, 655

\bibitem[Zamaninasab et al.(2014)]{zcst14}
Zamaninasab M., Clausen-Brown E., Savolainen T., Tchekhovskoy A., 2014, Nature, 510, 126

\bibitem[Zhang et al.(1999)]{zea99}
Zhang Y.H., Celotti A., Treves A., Chiappetti L., et al., 1999, ApJ, 527, 719

\bibitem[Zdziarski \& B\"ottcher(2015)]{zb15}
Zdziarski A.A., B\"ottcher M., 2015, MNRAS, 450, L21

\end{thebibliography}
\end{document}